\documentclass[aps,prl,preprintnumbers,showpacs,nofootinbib,floatfix]{revtex4}
\usepackage{graphicx}% Include figure files
\usepackage{subfig}
\usepackage{comment}
\usepackage{dcolumn}% Align table columns on decimal point
\usepackage{bm}% bold math
\begin{document}
\newcommand{\newc}{\newcommand}
\newc{\ra}{\rightarrow}
\newc{\lra}{\leftrightarrow}
\newc{\beq}{\begin{equation}}
\newc{\eeq}{\end{equation}}
\newc{\barr}{\begin{eqnarray}}
\newc{\earr}{\end{eqnarray}}
%%%%%%%%%%%%%%%%%%%%%%%%%%%%%%%%%%%%%%%%%%%
\newcommand{\Od}{{\cal O}}
\newcommand{\lsim}   {\mathrel{\mathop{\kern 0pt \rlap
  {\raise.2ex\hbox{$<$}}}
  \lower.9ex\hbox{\kern-.190em $\sim$}}}
\newcommand{\gsim}   {\mathrel{\mathop{\kern 0pt \rlap
  {\raise.2ex\hbox{$>$}}}
  \lower.9ex\hbox{\kern-.190em $\sim$}}}
%\preprint{APS/123-QED}

\title{THE DIURNAL VARIATION OF THE  WIMP DETECTION EVENT RATES
IN DIRECTIONAL EXPERIMENTS
 }

\author{J. D. Vergados$^{(1)}$\thanks{Vergados@cc.uoi.gr} and  Ch.C. Moustakidis$^{(2)}$\thanks{moustaki@auth.gr }}
\affiliation{$^{(1)}${\it Theoretical Physics Division, University
of Ioannina, Ioannina, Gr 451 10, Greece,}}
\affiliation{$^{(2)}${\it
 Department of Theoretical Physics, Aristotle University of
Thessaloniki, \\54124 Thessaloniki, Greece.}}
\begin{abstract}
The recent WMAP data have confirmed that exotic dark matter
together with the vacuum energy (cosmological constant) dominate
in the flat Universe.  Modern particle theories naturally provide viable cold dark matter
candidates with masses in the GeV-TeV region. Supersymmetry provides the lightest supersymmetric particle (LSP), theories in extra dimensions supply the lightest Kaluza-Klein particle (LKP) etc. The nature of dark matter can only be unraveled only by its direct detection in the laboratory. All such candidates will be called WIMPs (Weakly Interacting Massive Particles). In any case the direct dark matter search, which amounts to
 detecting the recoiling nucleus, following its collision with  WIMP, is central to
particle physics and cosmology.
In this work we briefly review the theoretical elements relevant to the direct dark matter detection experiments, paying particular attention to directional experiments. i.e experiments in which, not only the energy but the direction of the recoiling nucleus is observed. Since the direction of observation is fixed with respect the the earth, while the Earth is rotating around its axes, the directional experiment will sample different parts of the sky at different times during the day. So, since the event rates are different when looking at different parts of the sky, the observed signal in such experiments will exhibit very interesting and characteristic periodic diurnal variation.
\end{abstract}
\pacs{ 95.35.+d, 12.60.Jv}
 %%%%%%%%%%%%%%%%%%%%%%%%%%%%%%%%%%%%%%%%%%%%%%%%%%%%%%%%%%%%%%%%%%%%
\date{\today}
%%%%%%%%%%%%%%%%%%%%%%%%%%%%%%%%%%%%%%%%%%%%%%%%%%%%%%%%%%%%%%%%%%%%%
\maketitle
% \input{kappaphase}
%%%%%%%%%%%%%%%%%%%%%%%%%%%%%%%%%%%%%%%%%%%%%%%%%%%%%%%%%%%%%%%%%%%%%%%%%%%%%%%%%%%%%%%%%%%%%%%%%%%%%%%%%%%%%%%%%%%%%%%%%%
\section{Introduction}
The combined MAXIMA-1 \cite{MAXIMA-1}, BOOMERANG \cite{BOOMERANG},
DASI \cite{DASI} and COBE/DMR Cosmic Microwave Background (CMB)
observations \cite{COBE} imply that the Universe is flat
\cite{flat01}
%, $\Omega=1.11\pm0.07$
and that most of the matter in
the Universe is Dark \cite{SPERGEL}. i.e. exotic. These results have been confirmed and improved
by the recent WMAP data \cite{WMAP06}. Combining the
the data of these quite precise experiments one finds:
$$\Omega_b=0.0456 \pm 0.0015, \quad \Omega _{CDM}=0.228 \pm 0.013 , \quad \Omega_{\Lambda}= 0.726 \pm 0.015.$$
%$$\Omega_b=0.05, \Omega _{CDM}= 0.25, \Omega_{\Lambda}= 0.70$$
Since the non exotic component cannot exceed $40\%$ of the CDM
~\cite {Benne}, exotic (non baryonic) matter is required and there is room for cold dark matter candidates or WIMPs ( Weakly Interacting Massive Particles).
%  In fact the DAMA experiment ~\cite {BERNA2} has claimed the observation of one signal in direct
%detection of a WIMP, which with better statistics has subsequently
%been interpreted as a modulation signal \cite{BERNA1}.

Even though there exists firm indirect evidence for a halo of dark matter
in galaxies from the
observed rotational curves, see e.g the review \cite{UK01}, it is essential to directly
detect
% \cite{GOODWIT}-\cite{KVprd}
such matter.
Until dark matter is actually detected, we shall not be able to
exclude the possibility that the rotation curves result from a
modification of the laws of nature as we currently view them.  This makes it imperative that we
invest a
maximum effort in attempting to detect dark matter whenever it is
possible. Furthermore such a direct detection will also
unravel the nature of the constituents of dark matter.
The possibility of such detection, however, depends on the nature of the dark matter
constituents.

Supersymmetry naturally provides candidates for the dark matter constituents. In the most favored scenario of supersymmetry the
LSP (Lightest Supersymmetric Particle) can be simply described as a Majorana fermion, a linear
combination of the neutral components of the gauginos and
higgsinos \cite{ref2a,ref2b,ref2c,ref2,ELLROSZ,Gomez,ELLFOR}.
% \cite{ref2a},\cite{ref2b},\cite{ref2c},\cite{ref2},\cite{ELLROSZ},\cite{Gomez},\cite{ELLFOR}
%  \cite{GOODWIT}-\cite{ref2}.
In most calculations the
neutralino is assumed to be primarily a gaugino, usually a bino.
% \section{The Essential Theoretical Ingredients  of Direct Detection.}

Since the WIMP's are  expected to be very massive, $m_{WIMP} > 10$ GeV,  and
extremely non relativistic, with average kinetic energy $\langle T\rangle  \approx
50 \ {\rm KeV} (m_{WIMP}/ 100 \ {\rm GeV} )$, they are not likely to excite the nucleus.
So they can be directly detected mainly via the recoiling of a nucleus
(A,Z) in elastic scattering. The event rate for such a process can
be computed from the following ingredients:
%%%%%%%%%%%%%%%%%%%%%%%%%%%%%%%%%%%%%%%%%%%%%%%%%%%%%%%%%%%%%%%%%%%%%%%%%%%%%%

%%%%%%%%%%%%%%%%%%%%%%%%%%%%%%%%%%%%%%%%%%%%%%%%%%%%%%%%%%%%%%%%%%%%%%%%%%%%%%%
\begin{enumerate}
\item An effective Lagrangian at the elementary particle (quark)
level obtained in the framework of the prevailing particle theory. For supersymmetry
this is achieved as described,
e.g., in Refs.~\cite{ref2a,ref2b,ref2c,ref2,ELLROSZ,Gomez,ELLFOR,JDV96,JDV06a}.

\item A well defined procedure for transforming the amplitude
obtained using the previous effective Lagrangian from the quark to
the nucleon level, i.e. a quark model for the nucleon, see e.g.  \cite{JDV06a,Dree00,Dree,Chen}. This step
in SUSY models is not trivial, since the obtained results depend crucially on the
content of the nucleon in quarks other than u and d.
%This is
%particularly true for the scalar couplings, which are proportional
%to the quark masses~\cite{Dree}$-$\cite{Chen}, \cite{JDV06} as well as the
%isoscalar axial coupling \cite{JELLIS93,JDV06}.

\item Knowledge of the relevant nuclear matrix elements
\cite{Ress,DIVA00}, obtained with as reliable as possible many
body nuclear wave functions. Fortunately in the case of the scalar
coupling, which is viewed as the most important, the situation is
a bit simpler, since  then one  needs only the nuclear form
factor.

\item Knowledge of the WIMP density in our vicinity and its velocity distribution.\\
 We do not
know for sure what this distribution is, but  it is not expected to depend on the
nature of the WIMP.   In other words all WIMPs are expected to have the same velocity
distribution and matter density. The particle density, however, which enters the event rate, is expected
to be inversely proportional to the WIMP mass. In the present work we will consider
a Maxwellian distribution.
% differing in their characteristic velocities.
\end{enumerate}
%%%%%%%%%%
In the standard nuclear recoil experiments one has the problem that the reaction of interest does not have a characteristic feature to distinguish it
from the background. So for the expected low counting rates the background is
a formidable problem. Some special features of the LSP-nuclear interaction can be exploited to reduce the background problems. Such are:
\begin{itemize}
\item The modulation effect.\\
This yields a periodic signal due to the motion of the earth around the sun. Unfortunately this effect is small, $<2\%$ for most targets. Furthermore
one cannot exclude  backgrounds with a seasonal variation.
\item Transitions to excited states.\\
In this case one need not measure nuclear recoils, but the de-excitation $\gamma$ rays. This can happen only in vary special cases since the average WIMP energy is too low to excite the nucleus. It has, however, been found that in the special case of the target $^{127}$I such a process is feasible \cite{VQS04} with branching ratios around $5\%$.
\item Detection of electrons produced during the WIMP-nucleus collision.\\
It turns out, however, that this production peaks at very low energies. So only gaseous TPC detectors can reach the desired level of $100$ eV. In such a case the number of electrons detected may exceed the number of recoils for a target with high $Z$ \cite{VE05,MVE05}.
\item Detection of hard X-rays produced when the inner shell holes are filled.\\
  It has been found \cite{EMV05} that in the previous mechanism inner shell electrons can be ejected. These holes can be filled by the Auger process or X-ray emission. For a target like Xe these X-rays are in the $30$ keV region with the rate of about 0.1 per recoil for a WIMP mass of $100$ GeV.
\end{itemize}
In the present paper we will focus on the characteristic signatures of the WIMP nucleus
interaction, which will manifest themselves in directional recoil experiments, i.e. experiments
in which the direction of the recoiling nucleus is observed \cite{DRIFT,SHIMIZU03,KUDRY04,GREEN05,KRAUSS,KRAUSS01,JDV03,JDVSPIN04,VF07}. We will concentrate on the standard Maxwell-Boltzmann (M-B)
distribution for the WIMPs of our galaxy and we will not be concerned with other  distributions
\cite{VEROW06,JDV09,TETRVER06,VSH08},
even though some of them
 may yield stronger directional signals. Among those one should
mention the late infall of dark matter into the galaxy, i.e caustic rings
 \cite{SIKIVI1,SIKIVI2,Verg01,Green,Gelmini}, dark matter orbiting the
 Sun \cite{KRAUSS} and Sagittarius dark matter \cite{GREEN02}.

We will present our results in such
a fashion that they do not depend on the specific properties of the dark matter candidate, except that
the candidate is cold and massive, $m_{WIMP} > 10$ GeV. So, since the nucleon cross section will be treated as a free parameter, the only parameters which count is the reduced
mass, the nuclear form factor and the velocity distribution. So our results apply to all WIMPs.
  In a previous paper we have found that the observed rate is correlated
 with the direction of the sun's motion \cite{JDV03,JDVSPIN04,VF07}. On top of this one will observe a
relatively large time dependent variation
of the rate due to the motion of the earth. Those features cannot be masked by any known background. To fully exploit
these features the detector should be able to distinguish head from tail of the track, i.e. the
sense of direction of motion \cite{CKS-DS05}. An  analysis of the requirements imposed on the directional
detectors, focusing especially on energy threshold effects and minimization of background,
 has recently appeared \cite{GREEN06}.
The currently planned experiments, however, may not be able to distinguish the
two possible senses along the line of nuclear recoil  \cite{SPOONER.PC}. Such experiments cannot, e.g., measure the backward-forward asymmetry.

Previous calculations considered the possibility of the variation of the event rate as the direction of observation of the nuclear recoil is varied in the galactic coordinates. This expected asymmetry is due to the motion of the sun with respect to the center of the galaxy.  The apparatus will, of course, be oriented in a direction defined in the local frame, i.e.  the line of observation will point to a point in the sky  specified, in the equatorial system,
 by right ascension $\alpha$ and inclination $\delta$. Thus due to the rotation of the earth the directional experiment will probe different parts of the galaxy during the day. Hence we expect to observe a diurnal variation.
 In light of this possibility we will utilise our previous directional calculations \cite{JDV03,JDVSPIN04,VF07} to  explore, which characteristics, if any,
 of the previous results persist, as a diurnal variation.

 %%%%%%%%%%%%%%%%%%%%%%%%%%%%%%%%%%%%%%%%%%%
 \section{WIMP-nucleon cross section}
 %%%%%%%%%%%%%%%%%%%%%%%%%%%%%%%%%%%%%%%%%%%%

The WIMP nucleon differential cross section in the non relativistic limit is given by
\beq
{\rm d} \sigma=\frac{1}{\upsilon}|{\cal M}_n|^2 \frac{{\rm d}^3 {\bf p}^{'}}{(2 \pi)^3}  \frac{{\rm d}^3 {\bf q}}{(2 \pi)^3} (2 \pi)^3\delta\left ({\bf p}-{\bf p}^{'} -  {\bf q} \right ) 2 \pi~ \delta \left (\frac{p^2}{2 m_{\chi}} -\frac{(p^{'})^2}{2 m_{\chi}}-\frac{q^2}{2 m_p}  \right ),
\eeq
where ${\bf p},{\bf p}^{'}$  are the momenta of the initial, final  WIMP and  ${\bf q}$ the momentum transferred to the nucleon assumed to be initially at rest. Momentum and energy conservation are incorporated by suitable $\delta$ functions. $m_{\chi}$ and $m_p$ are the masses of the WIMP and the nucleon respectively. ${\cal M}_n$ is the invariant amplitude at the nucleon level.\\
Integrating over the final WIMP momentum we get
\beq
{\rm d} \sigma=\frac{1}{\upsilon}\frac{1}{(2 \pi)^2}|{\cal M}_n|^2\delta \left (\frac{p^2}{2 m_{\chi}} -\frac{({\bf p}-{\bf q})^2)^2}{2 m_{\chi}}-\frac{q^2}{2 m_p}  \right ){\rm d}^3{\bf q}=\frac{1}{\upsilon}\frac{1}{2 \pi)}|{\cal M}_n|^2\delta \left (q \upsilon \xi -\frac{q^2}{2 \mu_p}  \right )q^2 {\rm d}q {\rm d}\xi,
\eeq
where $\upsilon$ is the WIMP velocity, $\mu_p \approx m_p$ is the reduced mass of the WIMP nucleon system and $\xi={\hat p}.{\hat q}$. The integration over $q$ is trivial due to the $\delta$ function. The allowed value of $q$ is $q=2 \mu_p \upsilon \xi$.
We thus get
\beq
{\rm d} \sigma=\frac{1}{\upsilon}\frac{1}{2 \pi}|{\cal M}_n|^2\frac{\left (2 \mu_p \upsilon \xi \right )^2}{\upsilon \xi}{\rm d} \xi=\frac{1}{2 \pi}|{\cal M}_n|^2 4 \mu_p^2 \xi {\rm d} \xi.
\eeq
The scattering is forward, $0 \le \xi \le  1$. We thus find the nucleon cross section to be
\beq
\sigma_n=\frac{1}{2 \pi} |{\cal M}_n|^2 2 \mu_p^2.
\eeq
In the present calculation we are not going to be concerned with calculating the nucleon cross section, which is perhaps the most important ingredient as was discussed in the introduction~\cite{ref2a,ref2b,ref2c,ref2,ELLROSZ,Gomez,ELLFOR,JDV96,JDV06a,Dree00,Dree,Chen}. We will instead treat it as an input parameter (see below).

%%%%%%%%%%%%%%%%%%%%%%%%%%%%%%%%%%%%%%%%%%%%%
\section{WIMP-nucleus cross section}
%%%%%%%%%%%%%%%%%%%%%%%%%%%%%%%%%%%%%%%%%%%%%%
Proceeding as above we obtain
\beq
{\rm d} \sigma=\frac{1}{2 \pi}|{\cal M}_A|^2 2 \mu_r^2 \xi {\rm d} \xi,
\eeq
%\beq
%d \sigma=\frac{1}{\upsilon}\frac{1}{2 \pi}|{\cal M_n}|^2\frac{\left (2 \mu_r \upsilon \xi \right )^2}{\upsilon \xi}d \xi
%\eeq
where now ${\cal M}_A$ is the corresponding amplitude at the nuclear level and $\mu_r$ is the WIMP-nucleus reduced mass.\\
In this case instead of the variable $\xi$ it is advantageous to use the energy transfer $Q=q^2/(2 m_A)$.  We prefer to use the dimensionless variable $u=(1/2) (qb)^2$, where $b$ is the harmonic oscillator size parameter of the shell model of the nucleus. Thus
\beq
u=\frac{1}{2} (q b)^2= 2 \mu_r^2 \upsilon^2 \xi^2\Rightarrow\xi=\frac{\sqrt{u}}{\sqrt{2} mu_r \upsilon b}.
\eeq
Using the standard formula for the size parameter we can express u in terms of the energy transfer Q:
\beq
u=\frac{Q}{Q_0}~,~Q_0=(b^2 A m_p)^{-1}=40 A^{-4/3} \mbox { MeV },
\eeq
we thus get
\beq
{\rm d} \sigma=\frac{1}{2 \pi}|{\cal M}_A|^2  \mu_r^2 \frac{{\rm d} u}{ (\mu_r b \upsilon)^2}.
\eeq
To proceed further we should relate $|{\cal M}_A|^2 $ to the more elementary amplitude $|{\cal M}_n|^2$. This of, course, depends on the model. In the usual scace of the coherent scattering we get
\beq
|{\cal M}_A|^2=A^2 F^2(u) |{\cal M}_n|^2,
\eeq
where $F(u)$ is the nuclear form factor. We can thus relate the cross sections
\beq
{\rm d} \sigma=A^2 F^2(u) \sigma_n \left ( \frac{\mu_r}{\mu_p} \right )^2\frac{{\rm d} u}{2 (\mu_r b \upsilon)^2}.
\eeq

%%%%%%%%%%%%%%%%%%%%%%%%%%%%%%%%%%%%%%%%%%%%
\section{Standard (non Directional) Rates}
%%%%%%%%%%%%%%%%%%%%%%%%%%%%%%%%%%%%%%%%%%%%
The event rate is for a given WIMP velocity is given  by
\beq
%\frac{d R}{d t}=\frac{\rho_{\chi}}{m_{\chi}}\frac{m_t}{A m_p}  A^2 F^2(u)
{\rm d} R=\frac{\rho_{\chi}}{m_{\chi}}\frac{m_t}{A m_p}  A^2 F^2(u)\sigma_n \left ( \frac{\mu_r}{\mu_p} \right )^2 \upsilon \frac{{\rm d} u}{2 (\mu_r b \upsilon)^2},
\eeq
where ${m_t}/{A m_p}$ is the number of nuclei in a target of mass $m_t$. Also
\beq
\frac{{\rm d} R}{ {\rm d}u}=\frac{\rho_{\chi}}{m_{\chi}} \frac{m_t}{A m_p}\sqrt{<\upsilon^2>} A^2  \sigma_n \left ( \frac{\mu_r}{\mu_p} \right )^2 \frac{{\rm d} t(u,\upsilon)}{{\rm d} u},
\label{Eq:dRdu}
\eeq
where
\beq
\frac{{\rm d} t(u,\upsilon)}{{\rm d} u}=\frac{\upsilon}{\sqrt{<\upsilon^2>}} F^2(u) \frac{1}{2 (\mu_r b \upsilon)^2}.
\label{Eq:dtdu}
\eeq
We must now fold ${\rm d} t(u,\upsilon)/{\rm d}u$ with the velocity distribution $f(y,\xi)$ in the local frame. We will assume that the velocity distribution is Maxwell-Boltzmann in the galactic frame, namely
\beq
f(y')=\frac{1}{ \pi \sqrt{\pi}} e^{-(y')^2}, \qquad y'=\frac{\upsilon'}{\upsilon_0},
\label{Eq:M-B}
\eeq
%We find it convenient to use the dimensionless variable $y=\upsilon/\upsilon_0$,
where $\upsilon'$ is the WIMP velocity in the galactic frame and $\upsilon_0$ is the sun's velocity with respect to the center of the galaxy. We have, however, to transform the distribution it to the local frame:
\beq
{\bf y}'={\bf y}+{\hat z}+\delta \left (\sin{\alpha}{\hat x}-\cos{\alpha}\cos{\gamma}{\hat y}+\cos{\alpha}\sin{\gamma} {\hat z}\right ),
\eeq
where $\delta=\upsilon_1/\upsilon_0=0.136$ with $\upsilon_1$ the velocity of the earth around the sun, $\alpha$ the phase of the earth ($\alpha$=0 on June 3nd) and $\gamma\approx\pi/6$.

%%%%%%%%%%%%%%%%%%%%%%%%%%%%%%%%%%%%%%%%%%%%%%%
\subsection{ No modulation ($\delta=0$)}
\begin{figure}
\begin{center}
%\rotatebox{90}{\hspace{0.0cm} $\sigma_p\rightarrow 10^{-5}$pb}
\includegraphics[height=.3\textheight]{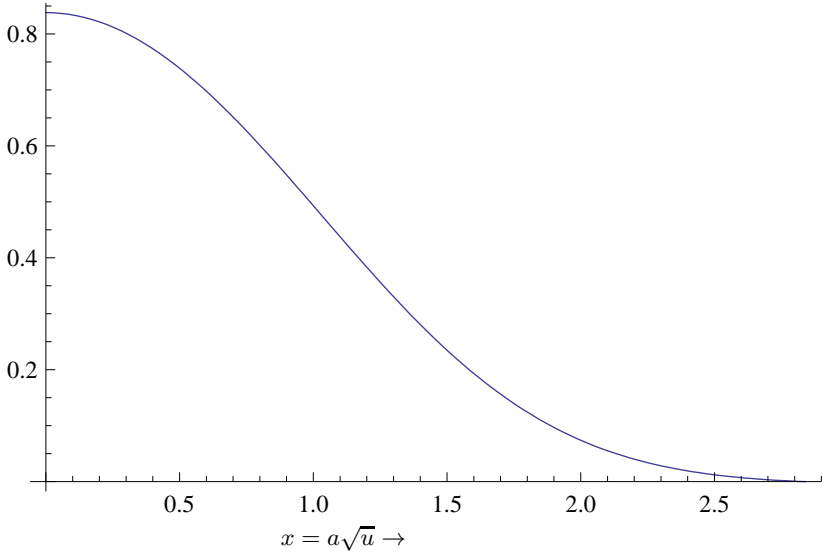}\\
\hspace{-2.0cm} $x=a\sqrt{u}\rightarrow$
\caption{ The function $\Psi_0(x)$ entering   the differential rate, which characterizes the velocity distribution.
 \label{fig:psi0}}
\end{center}
\end{figure}

%%%%%%%%%%%%%%%%%%%%%%%%%%%%%%%%%%%%%%%%%%%%%%%
With $\delta=0$ we find the distribution
\beq
f(y,\xi)=\frac{1}{\pi \sqrt{\pi}}e^{-(y^2+2 y \xi +1)}.
\eeq
%Assuming that the velocity distribution with respect to the center of the galaxy ,
Thus we get for the differential (with respect to the energy transfer) rate
\beq
\frac{{\rm d} R}{ {\rm d} u}=\frac{\rho_{\chi}}{m_{\chi}}\frac{m_t}{A m_p} \left ( \frac{\mu_r}{\mu_p} \right )^2 A^2\frac{d t}{du},\qquad \frac{{\rm d} t}{{\rm d} u}=\sqrt{\frac{2}{3}} a^2 F^2(u) \Psi_0(a \sqrt{u}),
\eeq
\beq
 \Psi_0(a \sqrt{u})=\int_{a\sqrt{u}} ^{y_{esc}} y {\rm d}y~ 2 \pi \int_{-1}^{\xi_0} {\rm d} \xi f(y,\xi),
\eeq
where $\xi_0 $ is a function of $y$ and may come from the requirement that the WIMP velocity must not exceed the escape velocity. In practice $\xi_0 \approx 1$ and also
\beq
a=(\sqrt{2} \mu_r b \upsilon_0)^{-1}.
\eeq
 The limits in $y$ are obtained as follows:
\beq y_{min}(u)\Leftrightarrow \xi=1\Rightarrow y_{min}(u)=a \sqrt{u}~,~y_{max}=y_{esc}=2.84,
\eeq
where the escape velocity is $\upsilon_{esc}=2.84 \upsilon_0$.\\
We note that the function $\Psi_0(x)$ depends only on the velocity distribution and is independent of nuclear physics (for $\xi_0=1$ will be given below, Eq.~(\ref{Eq:psi0})). It is a deceasing function of $u$ as shown in Fig.~\ref{fig:psi0}
%%%%%%%%%%%%%%%%%%%%%%%%%%%%%%%%%%%%%%%%%%%%%%%%%%%%%%%%%%%%%%%%%%%%%%%%%%%%%%%%%
%%%%%%%%%%%%%%%%%%%%%%%%%%%%%%%%%%%%%%%%%%%%%%%%%%%%%%%%%%%%%%%%%%%%%%%%%%%%%%%%%
 An additional suppression as the energy transfer increases comes, of course, from the nuclear form factor $ F(u)$. Note that the nuclear dependence of the differential rate comes not only from the form factor, but via the parameter $a$ as well.
Integrating the differential event rate from $u_{min}=E_{th}/Q_0$, which depends on the energy threshold, to $ u_{max}=y^2_{esc}/a^2$ we obtain the total rate
\beq
R=\frac{\rho_{\chi}}{m_{\chi}}\frac{m_t}{A m_p} \left ( \frac{\mu_r}{\mu_p} \right )^2 A^2t, \qquad t=\int_{u_{min}} ^{u_{max}} {\rm d} u\frac{{\rm d} t}{{\rm d} u}.
\eeq

%%%%%%%%%%%%%%%%%%%%%%%%%%%%%%%%%%%%%
\subsection{The modulation effect}
%%%%%%%%%%%%%%%%%%%%%%%%%%%%%%%%%%%%%

\begin{figure}
\begin{center}
%\rotatebox{90}{\hspace{0.0cm} $\sigma_p\rightarrow 10^{-5}$pb}
\includegraphics[height=.3\textheight]{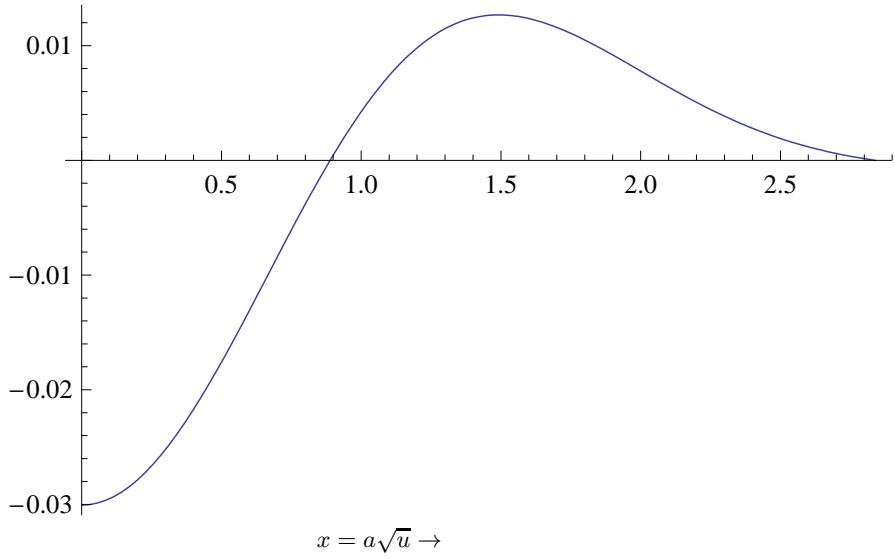}\\
\hspace{-2.0cm} $x=a\sqrt{u}\rightarrow$
\caption{ The function $\Psi_1(x,1/2,0.135)$ arising from the velocity distribution, which enters the modulated differential rate. Notice the change in sign. The position this happens depends on the reduced WIMP-nucleus mass. This dictates whether the modulation of the total rate is positive or negative.
 \label{fig:psi1}}
\end{center}
\end{figure}

Now the velocity distribution in the local frame becomes more complicated.
We get
\beq
f(y,\xi,\phi,\alpha,\gamma,\delta)=\frac{1}{\pi \sqrt{\pi}}e^{-(y^2+2 \xi y a+b+c)},
\eeq
with
\beq
A=\delta  \cos {\alpha } \sin {\gamma }+1~,~B=\delta ^2+2 \cos {\alpha } \sin {\gamma } \delta +1,
\eeq
and
\beq
C=2 y \delta  \sqrt{1-\xi ^2} \left (\sin{\alpha } \cos {\phi }-\cos
   {\alpha } \cos {\gamma } \sin {\phi } \right ).
\eeq
We now notice that to leading order in $\delta$ the term $C$ does not contribute, since the integration over $\phi$ yields zero. Then the integration over $\phi$ and $\xi$ can be done analytically to yield
\beq
f(y,\alpha,\gamma,\delta)=\frac{2 e^{-y^2-\delta ^2-2 \delta  \cos {\alpha } \sin {\gamma }-1}
   \sinh \left(\frac{}{}2 (\delta  \cos {\alpha } \sin {\gamma } y+y)\right)}{\sqrt{\pi }
   (\delta  \cos {\alpha } \sin {\gamma } y+y)}.
\eeq
The integration over $y$: $$\Psi(a \sqrt{u},\alpha,\gamma,\delta,y_{esc})= \int_{a\sqrt{u}}^{y_{esc}}y {\rm d} yf(y,\alpha,\gamma,\delta),$$
can also be done analytically to yield
\barr
\Psi(x,\alpha,\gamma,\delta,y_{esc})&=&
\frac{e^{-\delta ^2-2 \cos {\alpha } \sin {\gamma } \delta +(\delta
   \cos {\alpha } \sin {\gamma }+1)^2-1}}{{2 (\delta\cos {\alpha } \sin {\gamma }+1)}}
\nonumber\\
 && \left[\frac{}{}\text{erf}(-x+\delta
   \cos {\alpha } \sin {\gamma }+1)+\text{erf}(x+\delta  \cos {\alpha
   } \sin {\gamma }+1) \right.
\nonumber\\
&-&\left. \text{erf}\left(\delta  \cos {\alpha } \sin
   {\gamma }-y_{\text{esc}}+1\right)-\text{erf}\left(\delta  \cos
   {\alpha } \sin {\gamma }+y_{\text{esc}}+1\right) \frac{}{}\right],
\earr
with $x=a \sqrt{u}$ and erf(t) the well known error function. Expanding this in a power series up to linear terms  in $\delta$ we get:
\beq
\Psi(x)=\Psi_0(x)+\Psi_1(x,\gamma,\delta) \cos{\alpha},
\eeq
\beq
\Psi_0(x)=\Psi(x,\alpha,\gamma,0,y_{esc})=\frac{1}{2} \left  (\frac{}{}   \text{erf}(1 - x) + \text{erf}(1 + x) - \text{erf}(1 - y_{esc}) -
   \text{erf}(1 + y_{esc})\right ),
\label{Eq:psi0}
\eeq
for the zeroth order term  and
\barr
\Psi_1(x,\gamma,\delta)&=&\frac{1}{2} \delta
   \left[\frac{}{}-\text{erf}(1-x)-\text{erf}(x+1)+\text{erf}\left(1-y_{\text{e
   sc}}\right)+\text{erf}\left(y_{\text{esc}}+1\right)+\frac{2
   e^{-(x-1)^2}}{\sqrt{\pi }} \right.
\nonumber\\
&+&\left. \frac{2 e^{-(x+1)^2}}{\sqrt{\pi
   }}-\frac{2 e^{-\left(y_{\text{esc}}-1\right){}^2}}{\sqrt{\pi
   }}-\frac{2 e^{-\left(y_{\text{esc}}+1\right){}^2}}{\sqrt{\pi
   }}\frac{}{}\right] \sin (\gamma ),
\earr
for the coefficient of $\cos{\alpha}$ of the first order term. This is exhibited in Fig. \ref{fig:psi1}.
%%%%%%%%%%%%%%%%%%%%%%%%%%%%%%%%%%%%%%%%%%%%%%%%%%%%%%%%%%%%%%%%%%%%%%%%%%%%%%%%%
%%%%%%%%%%%%%%%%%%%%%%%%%%%%%%%%%%%%%%%%%%%%%%%%%%%%%%%%%%%%%%%%%%%%%%%%%%%%%%%%%
We thus proceed as above with
\beq
\Psi_0(x)\rightarrow\Psi_0(x)+\Psi_0(x,\gamma,\delta) \cos{\alpha},
\eeq
\beq
\frac{{\rm d} t}{{\rm d} u}\rightarrow\frac{{\rm d} r}{{\rm d} u}=\frac{{\rm d} t}{{\rm d} u}+\frac{{\rm d} \tilde{h}}{{\rm d} u} \cos{\alpha},
\eeq
\beq
t\rightarrow t(1+h\cos{\alpha}), \quad h=\frac{1}{t}\int_{u_{min}}^{u_{max}} {\rm d}u \frac{{\rm d} \tilde{h}}{{\rm d} u}.
\eeq
Thus we finally get
\beq
R=\frac{\rho_{\chi}}{m_{\chi}}\frac{m_t}{A m_p} \left ( \frac{\mu_r}{\mu_p} \right )^2 A^2~t\left (1+h \cos{\alpha} \right ).
\label{Eq:totalR}
\eeq

%%%%%%%%%%%%%%%%%%%%%%%%%%%%%%
\section{Directional Rates}
%%%%%%%%%%%%%%%%%%%%%%%%%%%%%%
In this instance the experiments will attempt to measure not only the energy, but the direction of the recoiling nucleus as well \cite{VF07},  \cite{JDV04}.
Let us indicate the direction observation by:
\beq
{\hat e}=(e_x,e_y,e_z)=(\sin{\Theta}\cos{\Phi},\sin{\Theta}\sin{\Phi},\cos{\Theta}).
\eeq
Now (\ref{Eq:dtdu}) must be notified since the variables $u$ and $\upsilon$ are no longer independent. Furthermore no integration over the azimuthal angle specifying the direction of the outgoing nucleus is performed. Eqs.~(\ref{Eq:dRdu}) and (\ref{Eq:dtdu} ) now become
\beq
\left (\frac{{\rm d} R}{ {\rm d} u}\right )_{dir}=\frac{\rho_{\chi}}{m_{\chi}} \frac{m_t}{A m_p}\sqrt{<\upsilon^2>} A^2  \sigma_n \left ( \frac{\mu_r}{\mu_p} \right )^2 \frac{1}{2 \pi}\left (\frac{{\rm d} t(u,\upsilon)}{{\rm d} u} \right )_{dir},
\label{Eq:dirdRdu}
\eeq
\beq
\left (\frac{{\rm d} t(u,\upsilon)}{{\rm d} u}\right )_{dir}=\frac{\upsilon}{\sqrt{<\upsilon^2>}} F^2(u) \frac{1}{2 (\mu_r b \upsilon)^2}\delta\left ({\hat \upsilon}.{\hat e}-\frac{a \sqrt{u}}{\upsilon/\upsilon_0}\right ).
\label{Eq:dirdtdu}
\eeq
The factor $1/( 2 \pi) $ enters since we are going to make use of the same nuclear cross section as in the non directional case.\\
Because of this constraint and the fact that the limits of integration over the WIMP velocity depend on the angles $\Theta$ and $\Phi$. We thus find it convenient to use a more convenient system $(X,Y,Z)$, in which the polar axes ${\hat Z}$ is in the direction of the recoiling nucleus. The needed transformation is
\beq
 \left ( \begin {array} {l} e_x \\  e_y \\ e_z
     \end {array}
    \right )=\left (
\begin{array}{lll}
 \cos (\Theta ) \cos (\Phi ) & -\sin (\Phi ) & \cos (\Phi ) \sin
   (\Theta ) \\
 \cos (\Theta ) \sin (\Phi ) & \cos (\Phi ) & \sin (\Theta ) \sin (\Phi
   ) \\
 -\sin (\Theta ) & 0 & \cos (\Theta )
\end{array}
\right )
\left ( \begin {array} {l} e_X \\  e_Y \\ e_Z
     \end {array}
\right ).
\eeq
and thus the three components of the velocity are given by:
\barr
& &\left ( \begin {array} {l} {\hat \upsilon} _X \\  {\hat \upsilon} _Y \\ {\hat \upsilon} _Z
     \end {array} \right )
\left (
\begin{array}{lll}
 \cos (\Theta ) \cos (\Phi ) & \cos (\Theta )
   \sin (\Phi ) & -\sin (\Theta ) \\
 -\sin (\Phi ) & \cos (\Phi ) & 0 \\
 \cos (\Phi ) \sin (\Theta ) & \sin (\Theta )
   \sin (\Phi ) & \cos (\Theta )
\end{array}
\right )
\left(
\begin{array}{l}
 \delta  \sin (\alpha ) \\
 -\delta  \cos (\alpha ) \cos (\gamma ) \\
 \delta  \cos (\alpha ) \sin (\gamma )+1
\end{array}
\right )
\\
&=&\left(
\begin{array}{l}
 \delta  \cos (\Theta ) \cos (\Phi ) \sin (\alpha
   )-(\delta  \cos (\alpha ) \sin (\gamma )+1)
   \sin (\Theta )-\delta  \cos (\alpha ) \cos
   (\gamma ) \cos (\Theta ) \sin (\Phi ) \\
 -\delta  \cos (\alpha ) \cos (\gamma ) \cos
   (\Phi )-\delta  \sin (\alpha ) \sin (\Phi ) \\
 \cos (\Theta ) (\delta  \cos (\alpha ) \sin
   (\gamma )+1)+\delta  \cos (\Phi ) \sin (\alpha
   ) \sin (\Theta )-\delta  \cos (\alpha ) \cos
   (\gamma ) \sin (\Theta ) \sin (\Phi )
\end{array}
\right).\nonumber
\earr
The exponential of the M-B distribution ( Eq. (\ref{Eq:M-B})) in this new  local frame becomes
\barr
&&y^2+2y \left[\frac{}{}-\delta  \sqrt{1-\xi ^2} \sin (\phi ) \left(\frac{}{}\cos
   (\alpha ) \cos (\gamma ) \cos (\Phi )+\sin (\alpha )
   \sin (\Phi )\right)+\xi  \left(\frac{}{}\cos (\Theta ) (\delta  \cos (\alpha
   ) \sin (\gamma )+1)\right. \right. \nonumber\\
&&+\left. \left. \delta  \sin (\Theta ) (\cos (\Phi )
   \sin (\alpha )-\cos (\alpha ) \cos (\gamma ) \sin (\Phi
   ))\frac{}{}\right)-\sqrt{1-\xi ^2} \cos (\phi ) \left( \frac{}{}(\delta  \cos (\alpha
   ) \sin (\gamma )+1)  \sin (\Theta ) \right. \right.\nonumber\\
&&+\left.\left.  \delta  \cos (\Theta
   ) (\cos (\alpha ) \cos (\gamma ) \sin (\Phi )-\cos (\Phi
   ) \sin (\alpha ))\frac{}{} \right)\frac{}{}\right] +\delta ^2+2 \delta  \cos
   (\alpha ) \sin (\gamma )+1,
\earr
where in this new frame the direction of the velocity is given by $(\theta,\phi)$ and  ${\hat \upsilon}.{\hat e}=\cos{\theta}=\xi$.\\
Expanding in powers of $\delta$ up to first order we get
\beq
f(y,\xi,\phi, \Theta, \Phi, \alpha, \gamma, \delta )=e^{-\left ( y^2+2y \xi  \cos (\Theta ) -2y \sqrt{1-\xi ^2} \cos (\phi ) \sin (\Theta ) +1 \right )}(1+pc0+ps0+pcc+pcs+psc+pss ),
\eeq
with
\beq
pc0=\delta  \left[ \frac{}{}-2 \sin (\gamma )-2 y \xi  \left(\frac{}{}\cos (\Theta ) \sin
   (\gamma )-\cos (\gamma ) \sin (\Theta ) \sin (\Phi )\right)  \right] \cos{\alpha},
\eeq
\beq
ps0=\delta\left(\frac{}{} -2 y  \xi  \cos (\Phi ) \sin (\Theta )\right ) \sin{\alpha},
\eeq
\beq
pcc=\delta \left [\frac{}{} 2 y   \sqrt{1-\xi ^2} \left(\frac{}{}\sin (\gamma ) \sin (\Theta
   )+\cos (\gamma ) \cos (\Theta ) \sin (\Phi )\right)\right ] \cos{\alpha} \cos{\phi},
\eeq
\beq
pcs=\delta \left ( 2 y  \sqrt{1-\xi ^2} \cos (\gamma ) \cos (\Phi ) \right )\cos{\alpha} \sin{\phi},
\eeq
\beq
psc=\delta \left(-2 y  \sqrt{1-\xi ^2} \cos (\Theta ) \cos (\Phi )  \right)\sin{\alpha} \cos{\phi},
\eeq
\beq
pss=\delta \left ( 2 y   \sqrt{1-\xi ^2} \sin (\Phi )  \right ) \sin{\alpha} \sin{\phi}.
\eeq
We now find that the integrals over $\phi$ can be done analytically
\beq
\int_0^{2 \pi} e^{a \cos{\phi}} {\rm d} \phi=\int_0^{2 \pi} e^{a \sin{\phi}} {\rm d} \phi=2 \pi I_0(a),
\eeq
\beq
\int_0^{2 \pi} e^{a \cos{\phi}}\cos{\phi} {\rm d} \phi=\int_0^{2 \pi} e^{a \sin{\phi}} \sin{\phi} {\rm d} \phi=2 \pi I_1(a),
\eeq
\beq
\int_0^{2 \pi} e^{a \cos{\phi}}\sin{\phi} {\rm d} \phi=\int_0^{2 \pi} e^{a \sin{\phi}} \cos{\phi} {\rm d} \phi=0.
\eeq
where $I_0(a)$ and $I_1(a)$ are the modified Bessel functions.
%%%%%%%%%%%%%%%%%%%%%%%%%%%%%%%%%%%%%%
\subsection{No modulation $\delta=0$}
\begin{figure}
\begin{center}
%\rotatebox{90}{\hspace{0.0cm} $\sigma_p\rightarrow 10^{-5}$pb}
\includegraphics[height=.3\textheight]{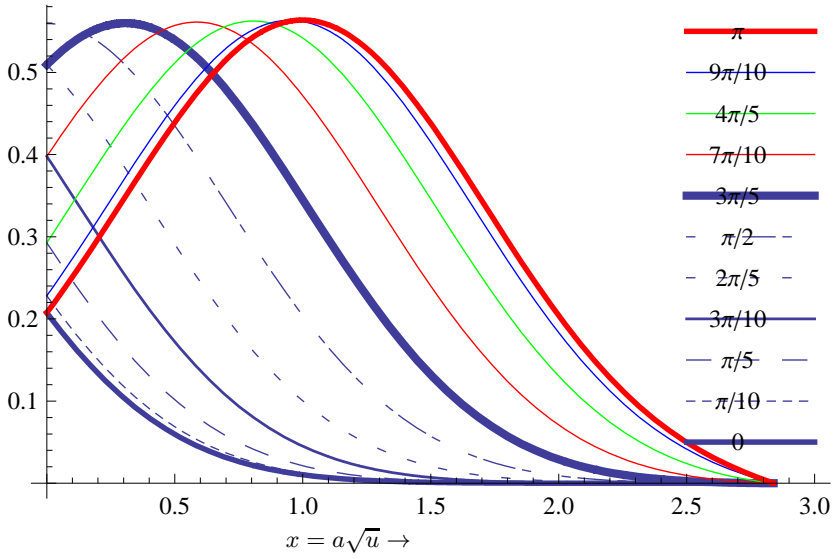}\\
\hspace{-2.0cm} $x=a\sqrt{u}\rightarrow$
\caption{ The function $\Psi^{dir}_0(x,\Theta)$  entering the directional differential rate.  Here we exhibit the curves corresponding $\Theta$ from zero to $\pi$ in steps of $\pi/10$ ( in the order from down up). We see that, with the possible exception of very small $x=a\sqrt{u}$ (small energy transfers) the differential rate increases with the angle $\Theta$ (maximum opposite to the sun's direction of motion).
 \label{fig:psi0dir}}
\end{center}
\end{figure}

%%%%%%%%%%%%%%%%%%%%%%%%%%%%%%%%%%%%%%
After the $\phi$ integration we get
\beq
f(y,\xi,\Theta)=\frac{2}{\sqrt{\pi}} e^{-y^2-2 \xi y  \cos ({\Theta }) -1} I_0\left (2 y
   \sqrt{1-\xi ^2} \sin (\Theta ) \right ).
\eeq
We can now use the $\delta$ function to perform the integral over $\xi$ to get:
\beq
f(x,y,\Theta)=\frac{2}{\sqrt{\pi}} e^{-y^2-2 x \cos ({\Theta }) -1}
I_0\left (2
   \sqrt{y^2-x^2} \sin (\Theta ) \right ),
\eeq
%\left[  I_0\left(
 %  \sqrt{1-\frac{x^2}{y^2}} \cos (\Phi ) \sin (\Theta
  % )\right)+  I_0\left(2 \sqrt{1-\frac{x^2}{y^2}} \sin
 %  (\Theta ) \sin (\Phi )\right)\right]
%\eeq
where $x=a\sqrt{u}$. The function $ \Psi^{dir}_0 (a \sqrt{u},\Theta,\Phi)$ can only be obtained numerically:
\beq
\Psi^{dir}_0 (a \sqrt{u},\Theta)=\int_{a \sqrt{u}}^{y_{esc}} y f(a \sqrt{u},y,\Theta) {\rm d}y.
\eeq
Once this function is known the computation of the event rate proceeds in a fashion analogous to the standard (non directional) case. The behavior of this function is shown in Fig. \ref{fig:psi0dir}.
%%%%%%%%%%%%%%%%%%%%%%%%%%%%%%%%%%%%%%%%%%%%%%%%%%%%%%%%%%%%%%%%%%%%%%%%%%%%%%%%%
%%%%%%%%%%%%%%%%%%%%%%%%%%%%%%%%%%%%%%%%%%%%%%%%%%%%%%%%%%%%%%%%%%%%%%%%%%%%%%%%%

%%%%%%%%%%%%%%%%%%%%%%%%%%%%%%%%%%%%%%%%%%%%%%%%%%%%%%%
\subsection{Modulated Rates in Directional experiments}
%%%%%%%%%%%%%%%%%%%%%%%%%%%%%%%%%%%%%%%%%%%%%%%%%%%%%%%

It is clear that the terms pcs and pss make no contribution. Thus after the $\phi$ integration we get
\beq
f(y,\xi, \Theta, \Phi, \alpha, \gamma, \delta )=
\frac{2}{\sqrt{\pi }} e^{-y^2-2 \xi y  \cos ({\Theta }) -1} \left ( qc0+ qs0+qcc+qss \right ),
\eeq
where
\beq
qc0=\delta \left [\frac{}{}-2 \sin (\gamma )-2 y \xi  \left( \frac{}{}\cos (\Theta ) \sin
   (\gamma )-\cos (\gamma ) \sin (\Theta ) \sin (\Phi )\right) \right ]  I_0\left (2y
   \sqrt{1-\xi ^2} \sin (\Theta ) \right )\cos{\alpha},
\eeq
\beq
qs0=\delta \left (\frac{}{} -2 y   \xi  \cos (\Phi ) \sin (\Theta ) \right ) I_0\left (2y
   \sqrt{1-\xi ^2} \sin (\Theta )\right )\sin{\alpha},
\eeq
\beq
qcc=\delta   \left[ \frac{}{}2 y  \sqrt{1-\xi ^2} \left(\frac{}{} \sin (\gamma ) \sin (\Theta
   )+\cos (\gamma ) \cos (\Theta ) \sin (\Phi )\right)\right]  I_1\left (2y
   \sqrt{1-\xi ^2} \sin (\Theta )\right )\cos{\alpha},
\eeq
\beq
qsc=\delta \left ( -2 y  \sqrt{1-\xi ^2} \cos (\Theta ) \cos (\Phi ) \right )  I_1\left (2y
   \sqrt{1-\xi ^2} \sin (\Theta )\right )\sin{\alpha}.
\eeq
Proceeding as above we can use the $\delta$ function to integrate over $\xi$ and get two modulation amplitudes:
\begin{itemize}
\item The amplitude that multiplies the $\cos{\alpha}$ term is:\\
\beq
f_c(x,y,\Theta,\Phi,\delta)=\frac{2}{\sqrt{\pi }} e^{-y^2-2 x \cos ({\Theta })-1}(rc0+rcc),
\eeq
with
\beq
rc0=\delta  \left[ \frac{}{}-2 \sin (\gamma )-2 x  \left(\frac{}{}\cos (\Theta ) \sin
   (\gamma )-\cos (\gamma ) \sin (\Theta ) \sin (\Phi )\right)\right]  I_0\left (2
   \sqrt{y^2-x^2} \sin (\Theta ) \right ),
\eeq
\beq
rcc=2 \delta  \sqrt{y^2-x^2} \left(\frac{}{}\sin (\gamma ) \sin (\Theta
   )+\cos (\gamma ) \cos (\Theta ) \sin (\Phi )\right)   I_1\left (2
   \sqrt{y^2-x^2} \sin (\Theta )\right ).
\eeq
%\barr
%&&f_c(x,y,\Theta,\Phi,\delta)=\frac{4}{\sqrt{\pi }} e^{-y^2-2 x \cos {\Theta %}-1} \nonumber\\
%&&\delta
%   \left (\sqrt{1-\frac{x^2}{y^2}} y I_1\left(2
%   \sqrt{1-\frac{x^2}{y^2}} \sin (\Theta )\right) (\sin
%   (\gamma ) \sin (\Theta )+\cos (\gamma ) \cos (\Theta )
%   \sin (\Phi ))\right )\nonumber\\
%&&+\delta \left (I_0\left(2 \sqrt{1-\frac{x^2}{y^2}} \sin
%   (\Theta )\right) (x \cos (\Theta ) \sin (\gamma )+\sin
%   (\gamma )-x \cos (\gamma ) \sin (\Theta ) \sin (\Phi
%   ))\right )
%\earr
Thus we numerically get
\beq
\Psi^{dir}_c (a \sqrt{u},\Theta,\Phi,\delta)=\int_{a \sqrt{u}}^{y_{esc}}~ y ~f_c(a \sqrt{u},y,\Theta,\Phi\delta) {\rm d}y.
\eeq
\item The amplitude that multiplies the $\sin{\alpha}$ term is of the form:\\
\beq
f_s(x,y,\Theta,\Phi,\delta)=\frac{2}{\sqrt{\pi }} e^{-y^2-2 x \cos ({\Theta })-1}(rs0+rsc),
\eeq
where
\beq
rs0=-2  \delta   x \left( \frac{}{}\cos (\Phi ) \sin (\Theta ) \right ) I_0\left (2
   \sqrt{y^2-x^2} \sin (\Theta )\right ),
\eeq
\beq
rsc=-2  \delta  \sqrt{y^2-x^2}\left( \frac{}{} \cos (\Theta ) \cos (\Phi ) \right )  I_1\left (2
   \sqrt{y^2-x ^2} \sin (\Theta )\right ).
\eeq
%\barr
%f_s(x,y,\Theta,\Phi,\delta)&=&\frac{4}{\sqrt{\pi}}e^{-y^2-2 x \cos (\Theta )-1} %\nonumber\\
%&&x \delta \left(
%   I_0\left (2 \sqrt{1-\frac{x^2}{y^2}} \sin (\Theta
%   )\right) \cos (\Phi ) \sin (\Theta )\right )\nonumber\\
%&-& y \delta \left ( \sqrt{1-\frac{x^2}{y^2}}  I_1\left(2
%   \sqrt{1-\frac{x^2}{y^2}} \sin (\Theta )\right) \cos
%   (\Theta ) \cos (\Phi )\right )
%\earr
Then finally:
\beq
\Psi^{dir}_s (a \sqrt{u},\Theta,\Phi,\delta)=\int_{a \sqrt{u}}^{y_{esc}} ~y~ f_s(a \sqrt{u},y,\Theta,\Phi,\delta) {\rm d}y.
\eeq
\end{itemize}
The differential rate is given:
\barr
\left (\frac{{\rm d}R}{{\rm d}u} \right )_{dir}&=&\frac{\rho_{\chi}}{m_{\chi}}\frac{m_t}{A m_p} \left ( \frac{\mu_r}{\mu_p} \right )^2 A^2 \sqrt{\frac{2}{3}} a^2 F^2(u)
\nonumber\\
&& \frac{1}{2\pi}\left (\frac{}{}\Psi^{dir}_0 (a \sqrt{u},\Theta,\Phi)+\Psi^{dir}_c (a \sqrt{u},\Theta,\Phi,\delta)\cos{\alpha}+\Psi^{dir}_s (a \sqrt{u},\Theta,\Phi,\delta)\sin{\alpha} \right ).
\earr

\begin{figure}
\begin{center}
%\rotatebox{90}{\hspace{0.0cm} $\sigma_p\rightarrow 10^{-5}$pb}
\includegraphics[height=.18\textheight]{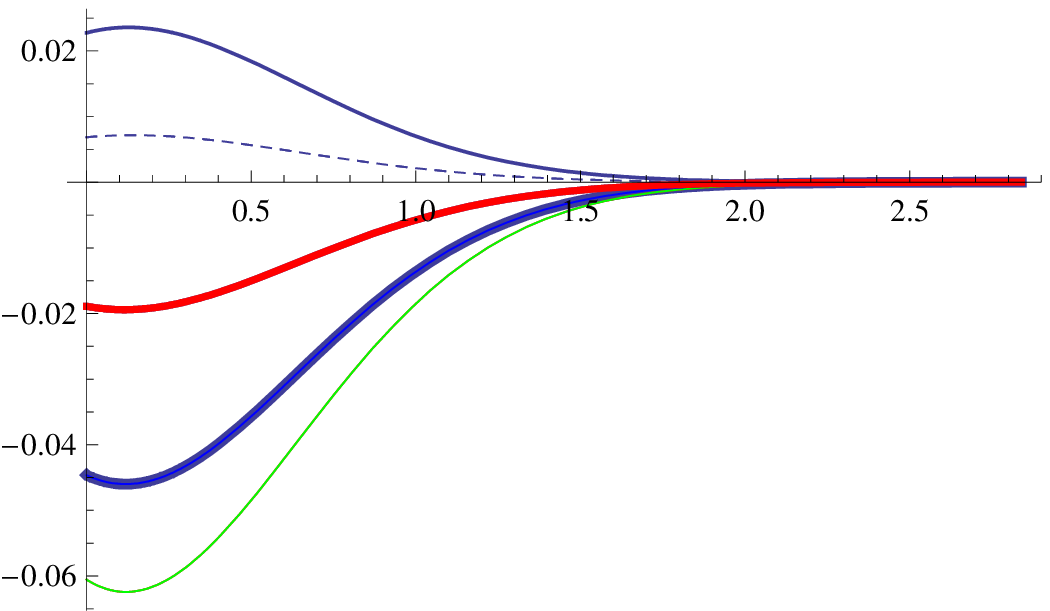}
\includegraphics[height=.18\textheight]{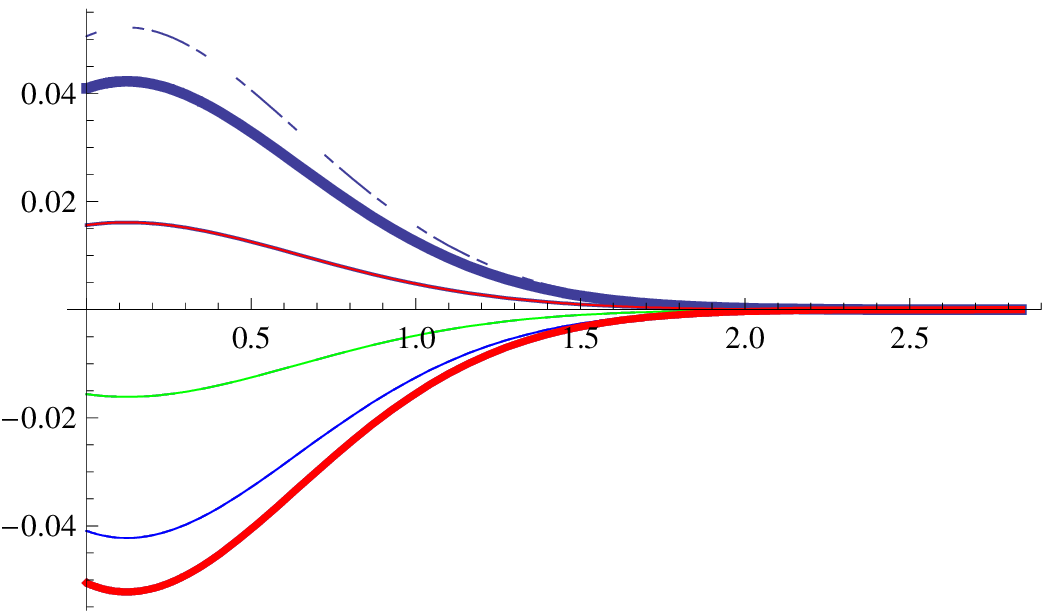}\\
\hspace{-2.0cm} $x=a\sqrt{u}\rightarrow$
\caption{ The function $\Psi^{dir}_c(x,\Theta,\Phi,\delta)$ on the left  and $\Psi^{dir}_s(x,\Theta,\Phi,\delta)$ entering the modulated directional differential rate as coefficients of the $\cos{\alpha}$ and  $\sin{\alpha}$ respectively . They correspond to $\Theta=(3/10)\pi$, i.e. almost parallel to the sun's direction of motion. The overall rate in this direction is expected to be small. The values of $\Phi$ range from zero to $2 \pi$ in steps of $\pi/5$ ( the style of the curves is the same as in Fig. \ref{fig:psi0dir}, but the step here is $\pi/5$). We see that the $\Phi$ dependence is strong at small energy transfers and but it gets smaller as the energy transfer increases.
 \label{fig:psi_sc_0.3pi}}
\end{center}
\end{figure}

\begin{figure}
\begin{center}
%\rotatebox{90}{\hspace{0.0cm} $\sigma_p\rightarrow 10^{-5}$pb}
\includegraphics[height=.18\textheight]{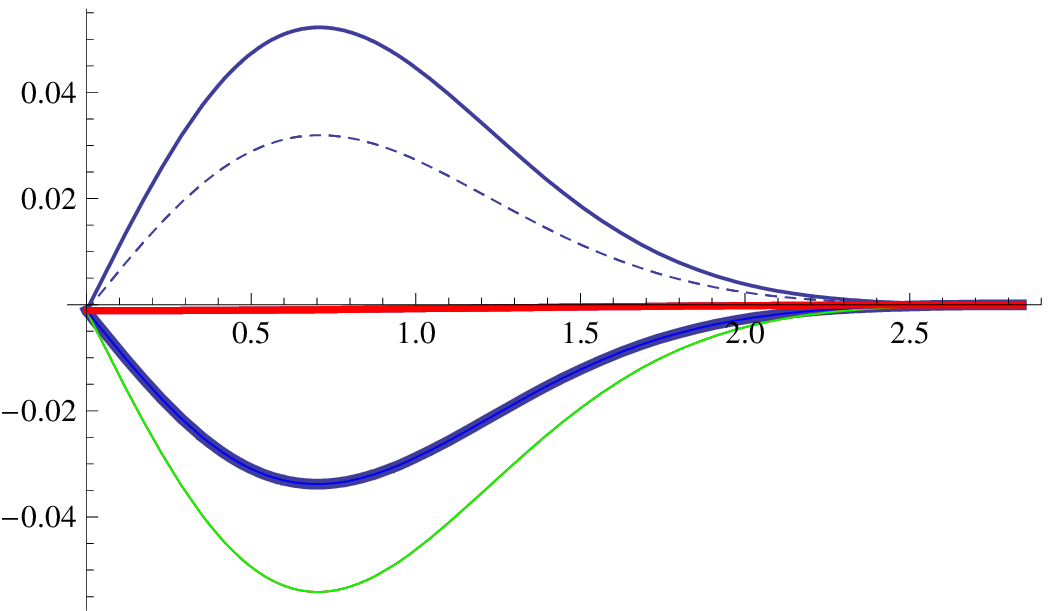}
\includegraphics[height=.18\textheight]{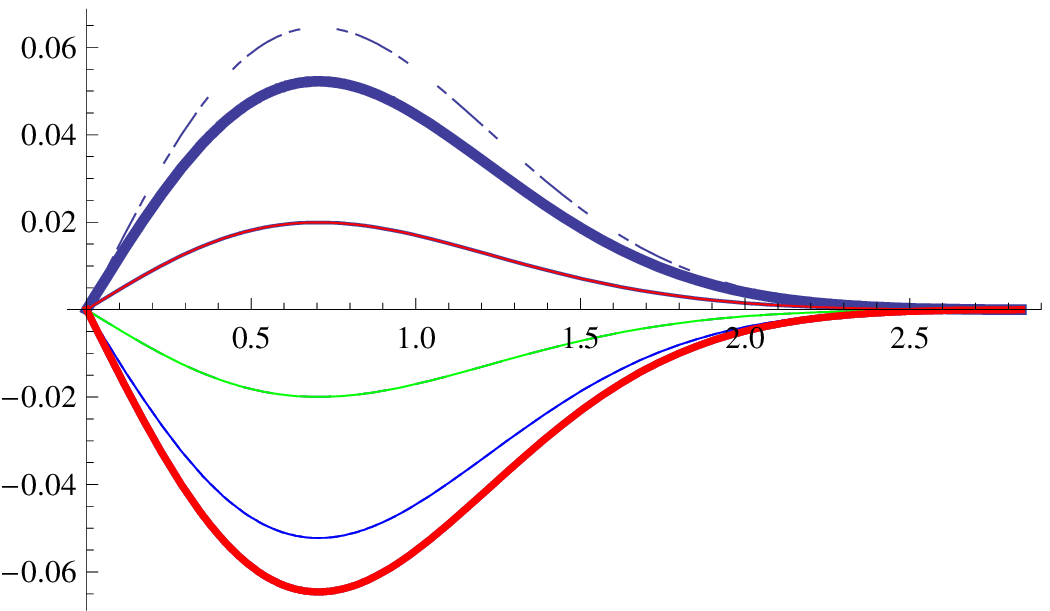}\\
\hspace{-2.0cm} $x=a\sqrt{u}\rightarrow$
\caption{ The same as in Fig. \ref{fig:psi_sc_0.3pi} for $\Theta=\pi/2$, i.e in a plane perpendicular to the sun's direction of motion.
 \label{fig:psi_sc_0.5pi}}
\end{center}
\end{figure}

\begin{figure}
\begin{center}
%\rotatebox{90}{\hspace{0.0cm} $\sigma_p\rightarrow 10^{-5}$pb}
\includegraphics[height=.18\textheight]{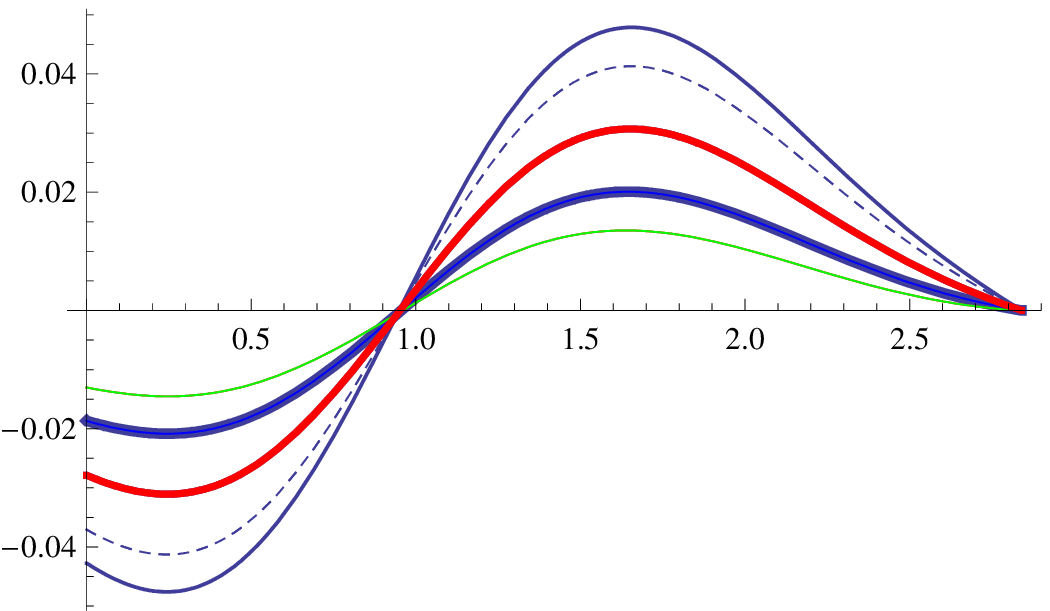}
\includegraphics[height=.18\textheight]{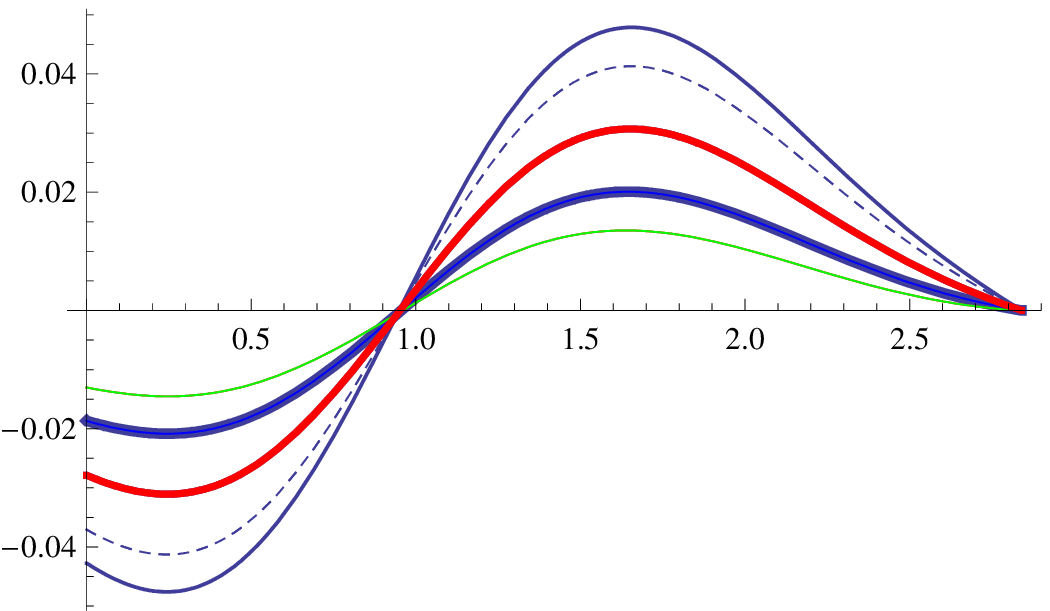}\\
\hspace{-2.0cm} $x=a\sqrt{u}\rightarrow$
\caption{ The same as in Fig. \ref{fig:psi_sc_0.3pi} for $\Theta=(9/10)\pi$, i.e very close to the direction opposite to that of the sun's velocity. In this case the two coefficients are almost identical and,  for given $\Phi$, the shape is similar to the non directional case (see Fig. \ref{fig:psi1}).
% The $\Phi$ dependence is not visible.
 \label{fig:psi_sc_0.9pi}}
\end{center}
\end{figure}

The behavior of the functions $\Psi^{dir}_c (x,\Theta,\Phi,\delta)$ and $\Psi^{dir}_s (a \sqrt{u},\Theta,\Phi,\delta)$, $x=a\sqrt{u}$, is exhibited in Figs \ref{fig:psi_sc_0.3pi} -\ref{fig:psi_sc_0.9pi}.
%%%%%%%%%%%%%%%%%%%%%%%%%%%%%%%%%%%%%%%%%%%%%%%%%%%%%%%%%%%%%%%%%%%%%%%%%%%%%%%%%
%%%%%%%%%%%%%%%%%%%%%%%%%%%%%%%%%%%%%%%%%%%%%%%%%%%%%%%%%%%%%%%%%%%%%%%%%%%%%%%%%
%%%%%%%%%%%%%%%%%%%%%%%%%%%%%%%%%%%%%%%%%%%%%%%%%%%%%%%%%%%%%%%%%%%%%%%%%%%%%%%%%
%%%%%%%%%%%%%%%%%%%%%%%%%%%%%%%%%%%%%%%%%%%%%%%%%%%%%%%%%%%%%%%%%%%%%%%%%%%%%%%%%
The total event rate is obtained after integrating  the above expression over the energy transfer. It can be cast in the form:
\beq
R_{dir}=\frac{\rho_{\chi}}{m_{\chi}}\frac{m_t}{A m_p} \left ( \frac{\mu_r}{\mu_p} \right )^2 A^2\frac{1}{2 \pi} t_{dir}(\Theta,\Phi)\left ( \frac{}{} 1+h_c(\Theta,\Phi) \cos{\alpha}+h_s(\Theta,\Phi) \sin {\alpha}\right ).
\eeq
We found it convenient to use a new  relative parameter  $\kappa$ (to avoid or minimize the dependence on the many parameters  of the particle model) as well as the parameters $h_m$ and $\alpha_0$ given by:
\beq
\kappa(\Theta,\Phi)=\frac{t_{dir}}{t},\qquad   h_m(\Theta,\Phi)\cos{(\alpha+\alpha_0)}=h_c(\Theta,\Phi) \cos{\alpha}+h_s(\Theta,\Phi) \sin {\alpha}.
\eeq

These parameters depend, of course, on the reduced mass $\mu_r$, the WIMP velocity and to some extend on the nuclear physics via the nuclear form factor.
The parameter $\kappa$ gives the retardation factor of the directional rate, over and above the factor of $1/(2 \pi)$, compared to the standard rate.  Since, however, the unmodulated amplitude is independent of $\Phi$ one can integrate over $\Phi$ so that the suppression of $1/(2 \pi)$ drops out for this term.
Because of the existence of both $\cos{\alpha}$ and $\sin{\alpha}$ terms the time dependence will be of the form $\cos{(\alpha+\alpha_0)}$, with the phase $\alpha_0 $ being direction dependent. Thus the time of the maximum and minimum will depend on the direction. In other words the seasonal dependence will depend on the direction of observation. So it cannot be masked by irrelevant seasonal effects.

Before concluding this section we like to consider  the case connected with the partly directional experiments, i.e. experiments which can detemine the line along which the nucleus is recoiling, but not the sense of direction on it. The results in this case can be obtained  by summing up the events in both directions. i.e. those specified by $(\Theta,\Phi)$ as well as  $(\pi-\Theta,\Phi+\pi)$. A given line of observation is now specified by $\Theta$, $\Phi$ in the range:
\[  0 \leq \Theta \leq \pi/2, \qquad   0 \leq \Phi \leq \pi. \]

%%%%%%%%%%%%%%%%%%%%%%%%%%%%%%%
 \section{Some Applications}
\begin{figure}[!ht]
 \begin{center}
  \includegraphics[scale=1.1]{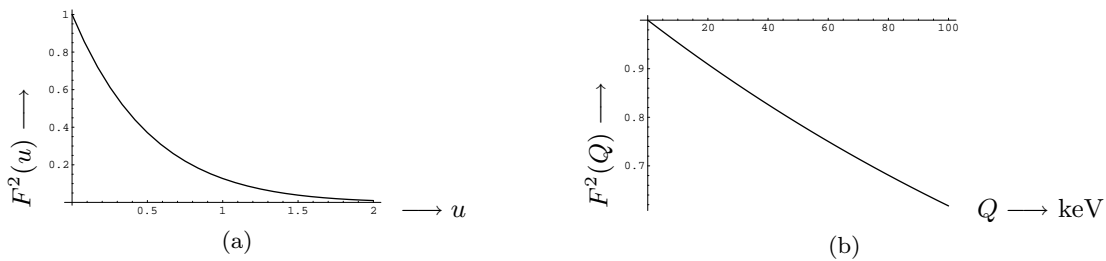}
   \caption{(a) The form factor $F^2(u)$ for $^{32}$S employed in our calculation with  $u=Q/Q_0$, Q  the
 energy transfer to the nucleus   and $Q_0=404$ keV. (b) The same quantity a function of the energy transfer $Q$.}
  \label{fig:sqformf32}
   \end{center}
  \end{figure}

\begin{figure}[!ht]
 \begin{center}
  \subfloat[]
 {
\rotatebox{90}{\hspace{-0.0cm} {$F^2(u) \longrightarrow$}}
\includegraphics[scale=0.45]{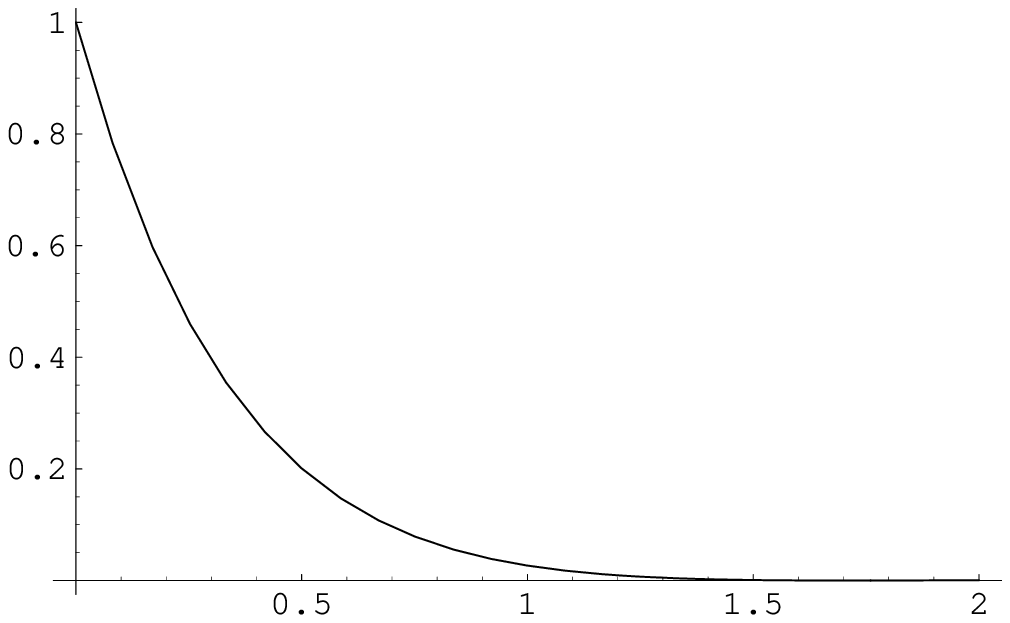}
{\hspace{0.0cm} $\longrightarrow  u$}
}
 \hspace{1.0cm}
 \subfloat[]
 {
\rotatebox{90}{\hspace{-0.0cm} {$F^2(Q) \longrightarrow$}}
\includegraphics[scale=0.45]{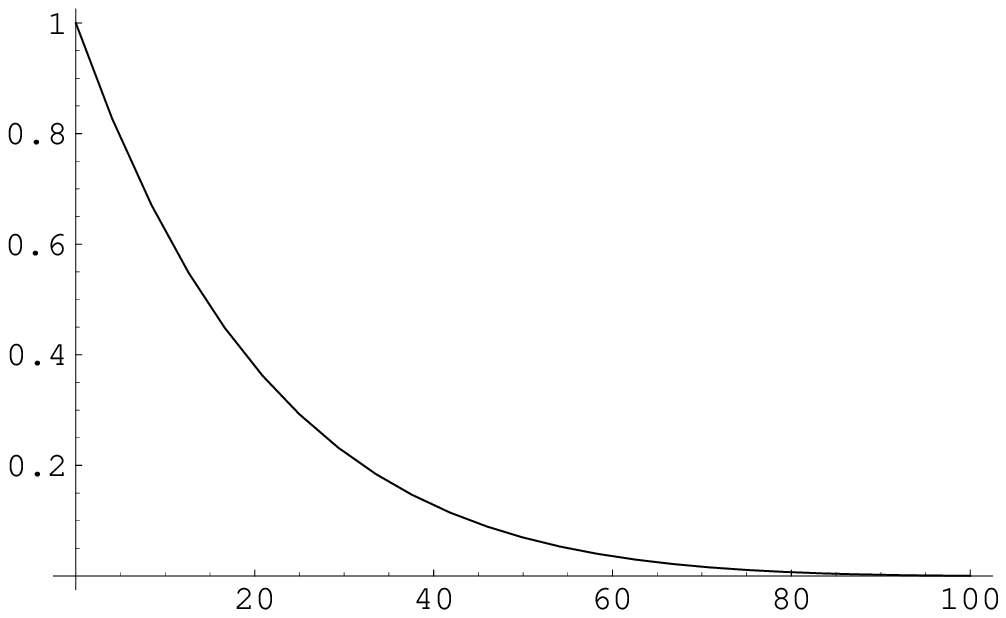}
{\hspace{0.0cm} $Q \longrightarrow  $ keV}
}
%{\hspace{-2.0cm}}
%{\hspace{-2.0cm}}
%{\hspace{4.0cm} $\longrightarrow  u=\frac{Q}{Q_0}$, $Q_0=64$ keV}
\caption{The same as in Fig. \ref{fig:sqformf32} in the case of a heavy target (A=127 or A=131).}
 \label{sqformf127}
   \end{center}
  \end{figure}

%%%%%%%%%%%%%%%%%%%%%%%%%%%%
In this section we are going to apply the formalism of the previous section in
 the case of two popular targets: i) The light target  $^{32}S$ appearing in CS$_2$ involved in DRIFT \cite{DRIFT}
  and ii) The heavy targets
$^{127}$I , which has been  employed in the DAMA experiment \cite{BERNA1,BERNA2} and $^{131}$Xe
\cite{XENON08,CDMSII04} employed by the XENON collaboration. We will not consider energy
thresh hold effects and we will
ignore quenching factor effects. We will consider only the coherent mode, but the results obtained for the functions
$t,\kappa$ are not expected to be radically modified, if one considers the spin mode (the quantities $h_m$ and $\alpha_0$ will be discussed in a future publication.
%\subsection{The light target CS$_2$}
The nuclear form factor employed was obtained in the shell model description of the target and is shown in
 Figs. \ref{fig:sqformf32} and \ref{sqformf127} for a light and a heavy target respectively.
 %%%%%%%%%%%%%%%%%%%%%%%%%%%%%%%%%%%%%%%%%%%%%%%%%%%%%%%%%%%%%%%%%%%%%%%%%%%%%%%%%
 %%%%%%%%%%%%%%%%%%%%%%%%%%%%%%%%%%%%%%%%%%%%%%%%%%%%%%%%%%%%%%%%%%%%%%%%%%%%%%%%%

 %%%%%%%%%%%%%%%%%%%%%%%%%%%%%%%%%%%%%%%%%%%%%%%%%%%%%%%%%%%%%%%%%%%%%%%%%%%%%%%%%
 %%%%%%%%%%%%%%%%%%%%%%%%%%%%%%%%%%%%%%%%%
  \subsection{The non directional case}
 %%%%%%%%%%%%%%%%%%%%%%%%%%%%%%%%%%%%%%%%%%%%%

\begin{figure}[!ht]
 \begin{center}
   \subfloat[]
 {
\rotatebox{90}{\hspace{-0.0cm} {$t_{coh} \longrightarrow$}}
\includegraphics[scale=0.6]{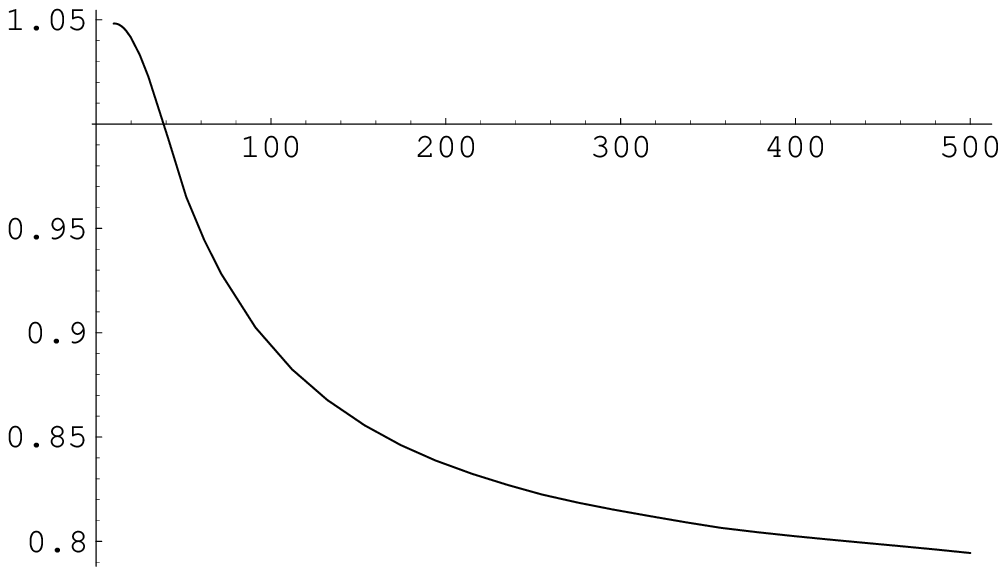}
}
  \subfloat[]
 {
\rotatebox{90}{\hspace{-0.0cm} {$t_{coh} \longrightarrow $}}
\includegraphics[scale=0.6]{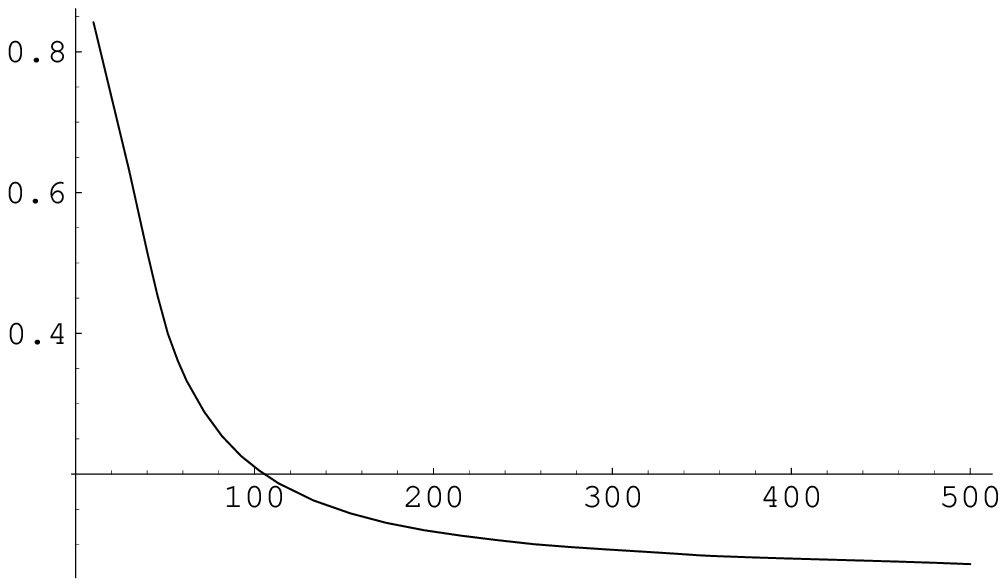}
}
\\
{\hspace{1.0cm} $m_{\chi}\longrightarrow $ GeV}
\caption{The quantity $t_{coh}$, as a function of the WIMP mass in GeV, is shown for  $Q_{min}=0$ for a light system, $^{32}$S (a) and  $^{127}$I or $^{131}$Xe (b).}
 \label{fig:totalt}
   \end{center}
   \end{figure}

\begin{figure}
   \begin{center}
   \subfloat[]
 {
\rotatebox{90}{\hspace{-0.0cm} {$h_{coh} \longrightarrow$}}
\includegraphics[scale=0.6]{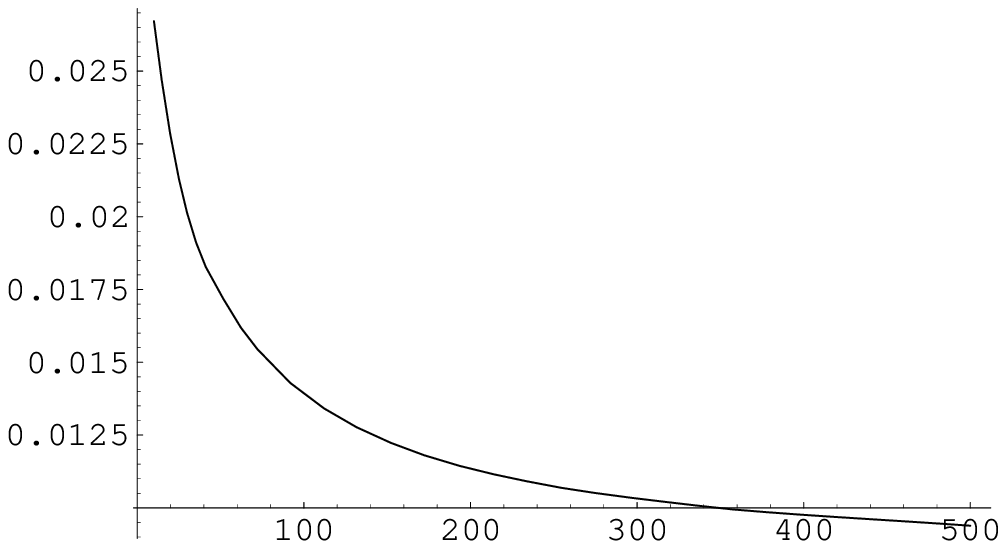}
}
  \subfloat[]
 {
\rotatebox{90}{\hspace{-0.0cm} {$h_{coh} \longrightarrow$}}
\includegraphics[scale=0.6]{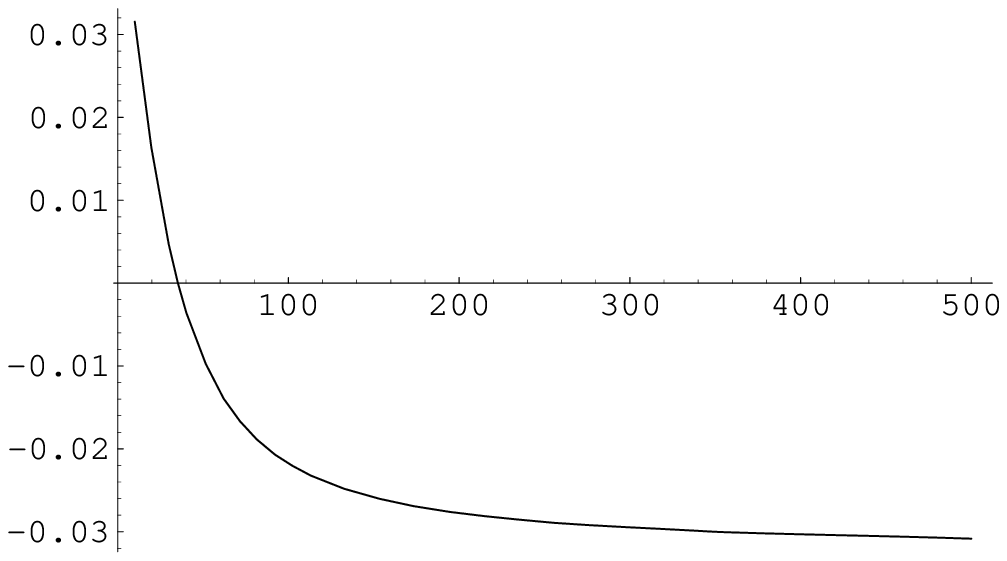}
}
\\
{\hspace{1.0cm} $m_{\chi}\longrightarrow $ GeV}
\caption{The quantity $h_{coh}$ is shown for  $Q_{min}=0$ for a light system, $^{32}$S (a) and  $^{127}$I or $^{131}$Xe (b).}
 \label{fig:totalh}
   \end{center}
  \end{figure}

\begin{figure}[!ht]
 \begin{center}
 \subfloat[]
 {
\rotatebox{90}{\hspace{-0.0cm} {$R \longrightarrow $ events/kg/y}}
\includegraphics[scale=0.6]{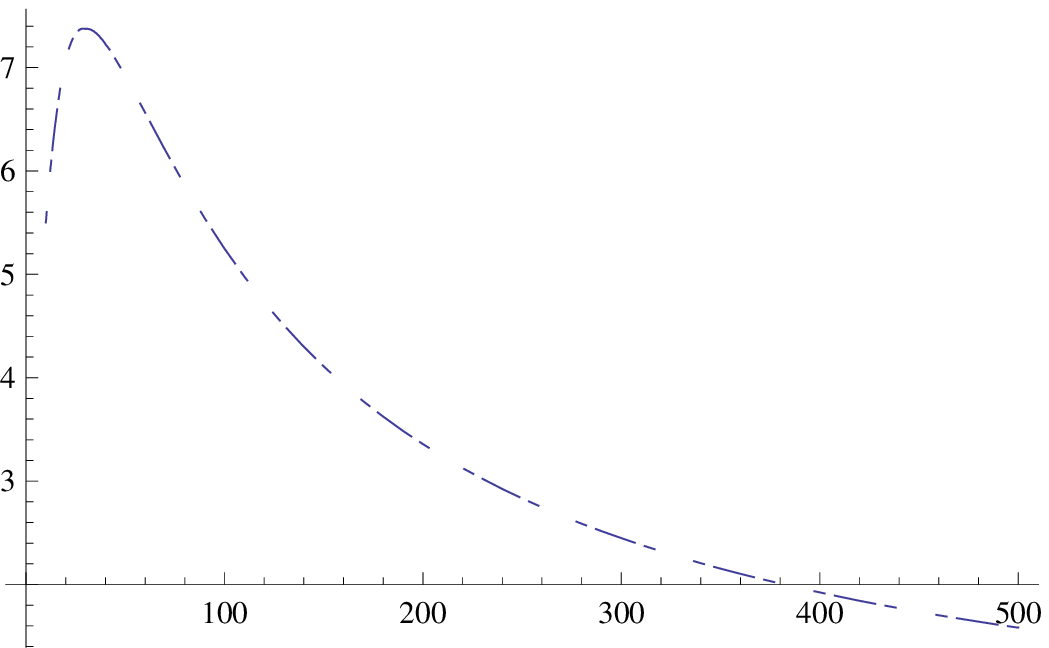}
}
  \subfloat[]
 {
\rotatebox{90}{\hspace{-0.0cm} {$R \longrightarrow $ events/kg/y}}
\includegraphics[scale=0.6]{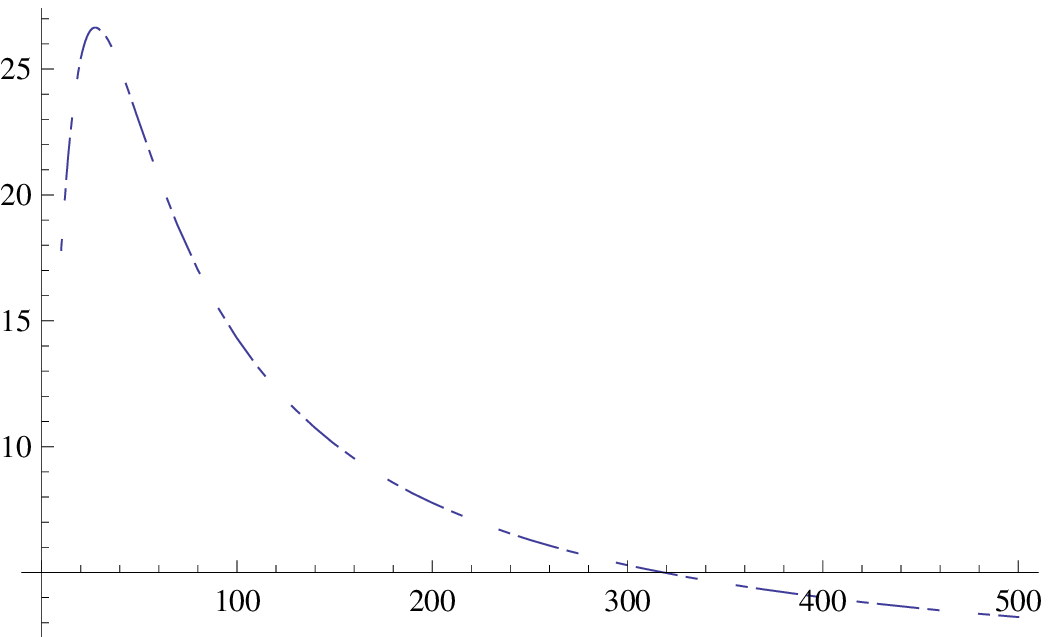}
}
\\
{\hspace{1.0cm} $m_{\chi}\longrightarrow $ GeV}
\caption{The coherent total event rate, as a function of the WIMP mass in GeV, is exhibited assuming a nucleon
 cross section $\sigma_N=10^{-32}$ cm$^2$
  for $Q_{min}=0$
   for a light system, $^{32}$S (a) and a heavy system,
     $^{131}$Xe (b).}
 \label{fig:totalR}
   \end{center}
   \end{figure}

  The parameter $t$ and $h$ entering the non directional case is shown in Figs. \ref{fig:totalt} and \ref{fig:totalh}.
%%%%%%%%%%%%%%%%%%%%%%%%%%%%%%%%%%%%%%%%%%%%%%%%%%%%%%%%%%%%%%%%%%%%%%%%%%%%%%%%%
%%%%%%%%%%%%%%%%%%%%%%%%%%%%%%%%%%%%%%%%%%%%%%%%%%%%%%%%%%%%%%%%%%%%%%%%%%%%%%%%%
%%%%%%%%%%%%%%%%%%%%%%%%%%%%%%%%%%%%%%%%%%%%%%%%%%%%%%%%%%%%%%%%%%%%%%%%%%%%%%%%%
 Using  the above parameters $t$ and employing  Eq. (\ref{Eq:totalR}) we obtain the time independent total
 event rate. The obtained results are shown in Fig. \ref{fig:totalR}. The total modulation, since it is
 defined relative to the time independent part, is still given by Fig. \ref{fig:totalh}.
 %%%%%%%%%%%%%%%%%%%%%%%%%%%%%%%%%%%%%%%%%%%%%%%%%%%%%%%%%%%%%%%%%%%%%%%%%%%%%%%%%
%%%%%%%%%%%%%%%%%%%%%%%%%%%%%%%%%%%%%%%%%%%%%%%%%%%%%%%%%%%%%%%%%%%%%%%%%%%%%%%%%
%%%%%%%%%%%%%%%%%%%%%%%%%%%%%%%%%%%%%
\subsection{The directional case}
\begin{figure}[!ht]
 \begin{center}
 \subfloat[]
 {
\rotatebox{90}{\hspace{-0.0cm}{$\kappa \longrightarrow$}}
\includegraphics[scale=0.5]{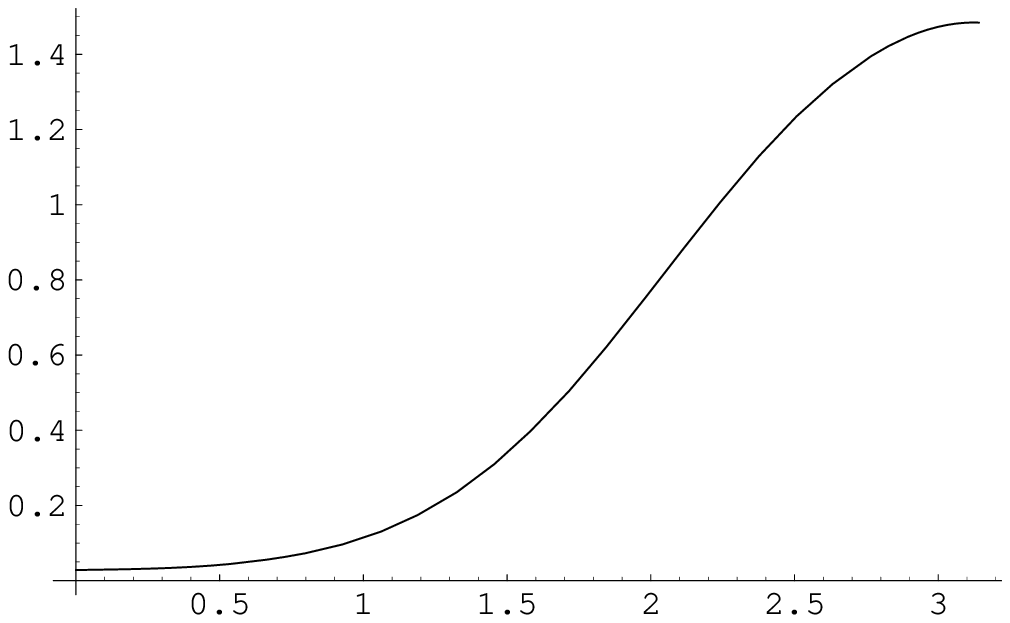}
}
 \hspace{1.0cm}
 \subfloat[]
 {
\rotatebox{90}{\hspace{-0.0cm} {$\kappa \longrightarrow$}}
\includegraphics[scale=0.5]{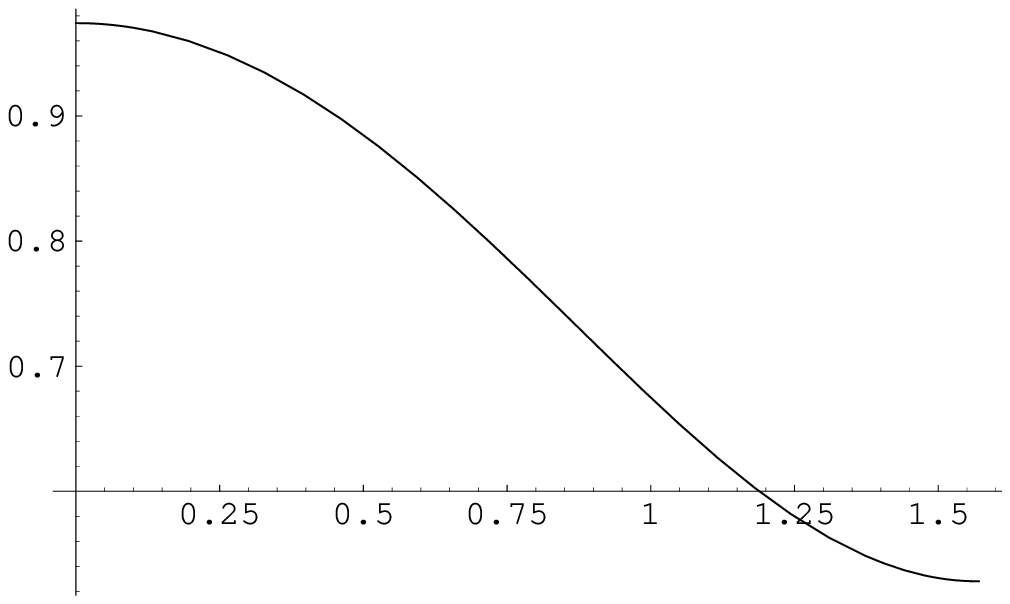}
}
\\
\hspace{-0.0cm} {$\Theta \longrightarrow$ radians}
\caption{The parameter $\kappa$ (see text) in the case of the target $^{32}$S for a WIMP mass of 1GeV, in the case the sense of direction  is known (a) and when the sense of direction cannot be distinguished (b).}
\label{fig:Skappa1}
  \end{center}
  \end{figure}

\begin{figure}[!ht]
 \begin{center}
 \subfloat[]
 {
\rotatebox{90}{\hspace{-0.0cm} {$\kappa \longrightarrow$}}
\includegraphics[scale=0.5]{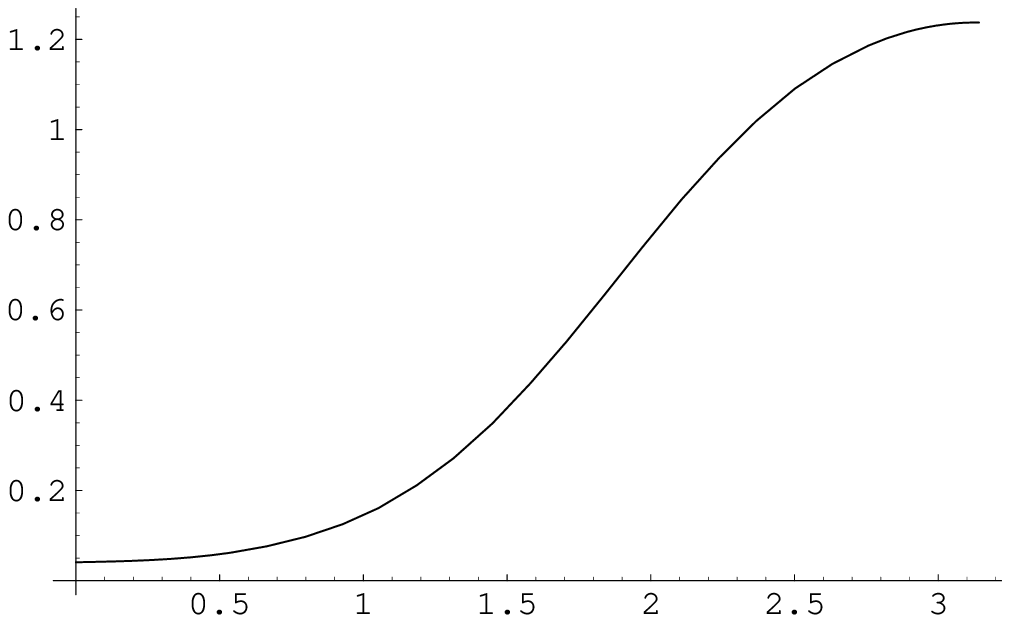}
}
 \hspace{1.0cm}
 \subfloat[]
 {
\rotatebox{90}{\hspace{-0.0cm} {$\kappa \longrightarrow$}}
\includegraphics[scale=0.5]{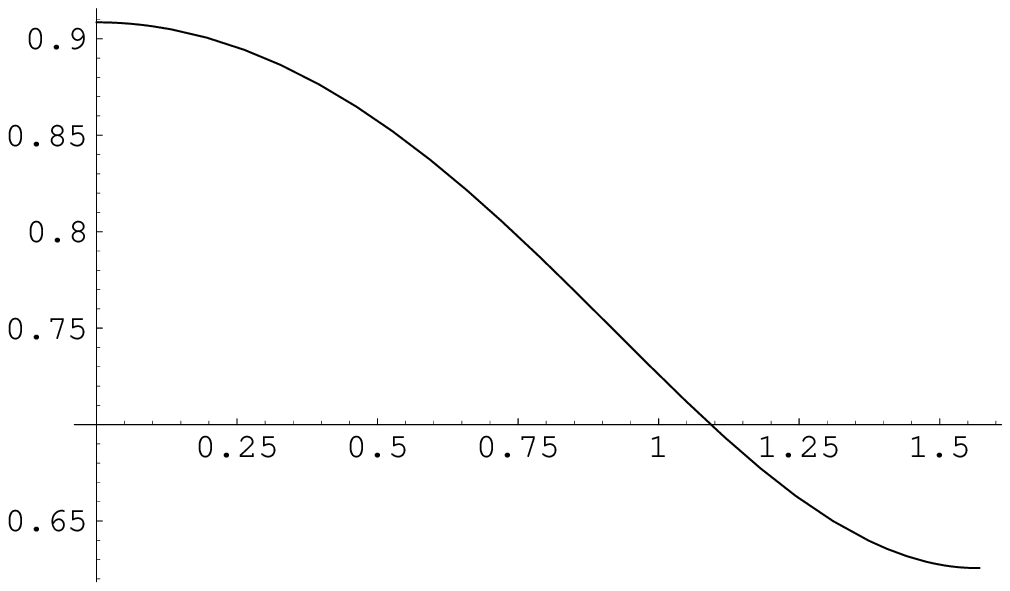}
}
\\
\hspace{-0.0cm} {$\Theta \longrightarrow$ radians}
\caption{The same as in Fig. \ref{fig:Skappa1} for a WIMP mass of 100 GeV.}
 \label{fig:Skappa2}
  \end{center}
  \end{figure}

\begin{figure}[!ht]
 \begin{center}
 \subfloat[]
 {
\rotatebox{90}{\hspace{-0.0cm} {$\kappa \longrightarrow$}}
\includegraphics[scale=0.5]{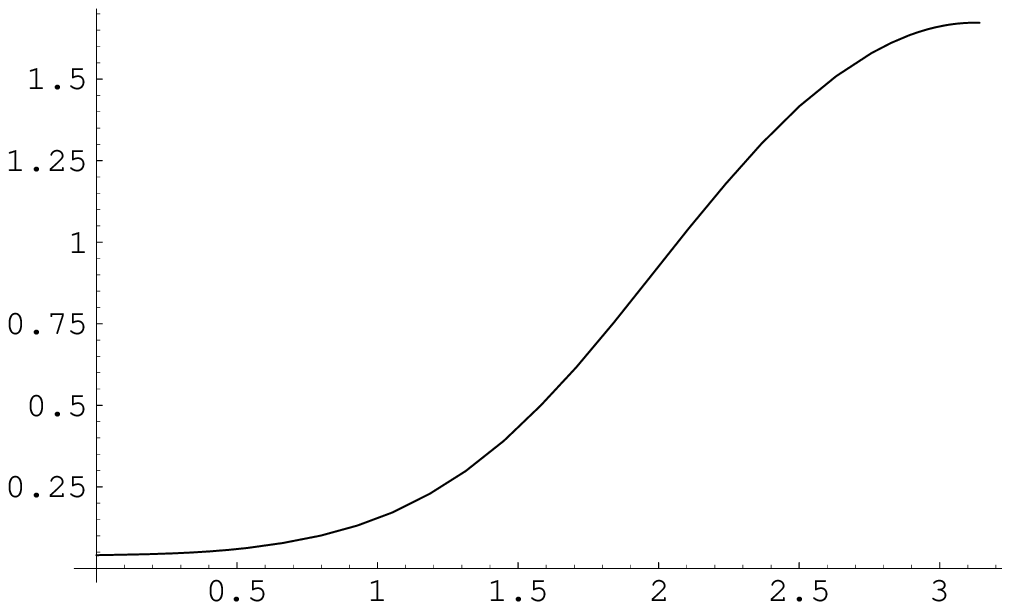}
}
 \hspace{1.0cm}
 \subfloat[]
 {
\rotatebox{90}{\hspace{-0.0cm} {$\kappa \longrightarrow$}}
\includegraphics[scale=0.5]{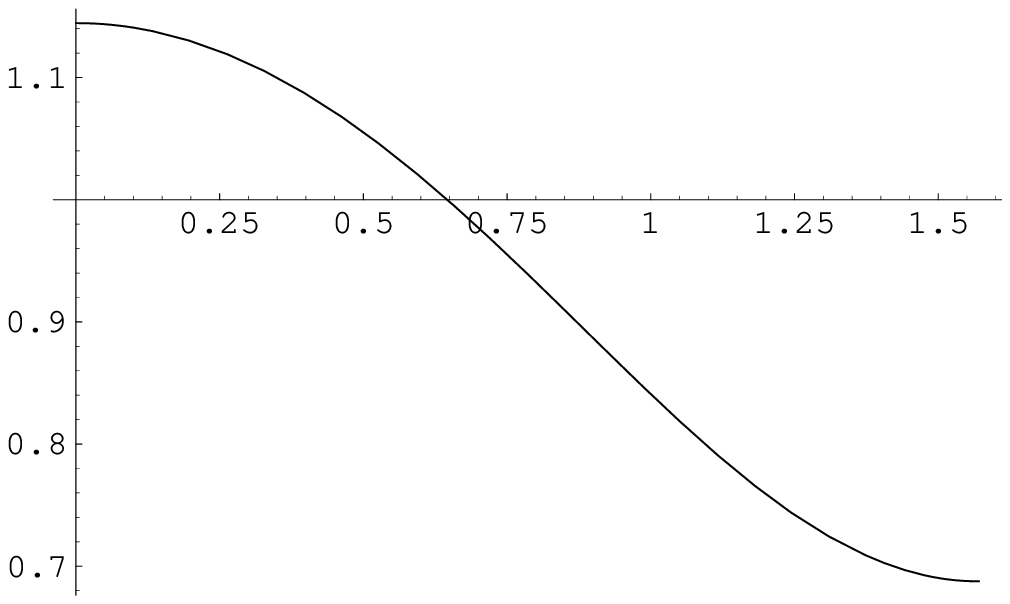}
}
\\
\hspace{-0.0cm} {$\Theta \longrightarrow$ radians}
\caption{The same as in Fig. \ref{fig:Skappa1} for a a heavy target.}
 \label{fig:kappa1}
  \end{center}
  \end{figure}

\begin{figure}[!ht]
 \begin{center}
 \subfloat[]
 {
\rotatebox{90}{\hspace{-0.0cm} {$\kappa \longrightarrow$}}
\includegraphics[scale=0.5]{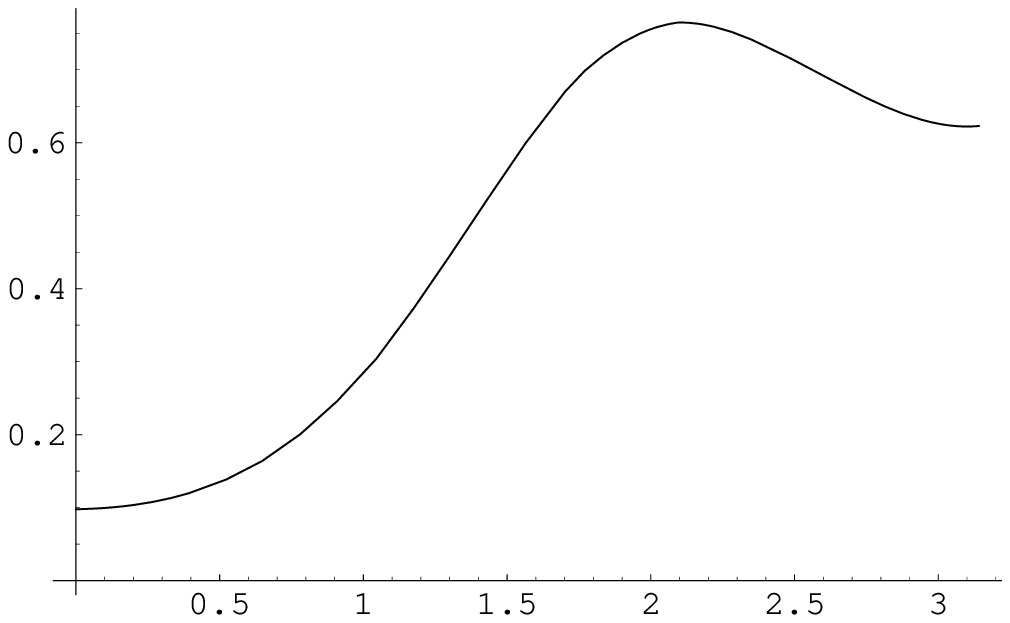}
}
 \hspace{1.0cm}
 \subfloat[]
 {
\rotatebox{90}{\hspace{-0.0cm} {$\kappa \longrightarrow$}}
\includegraphics[scale=0.5]{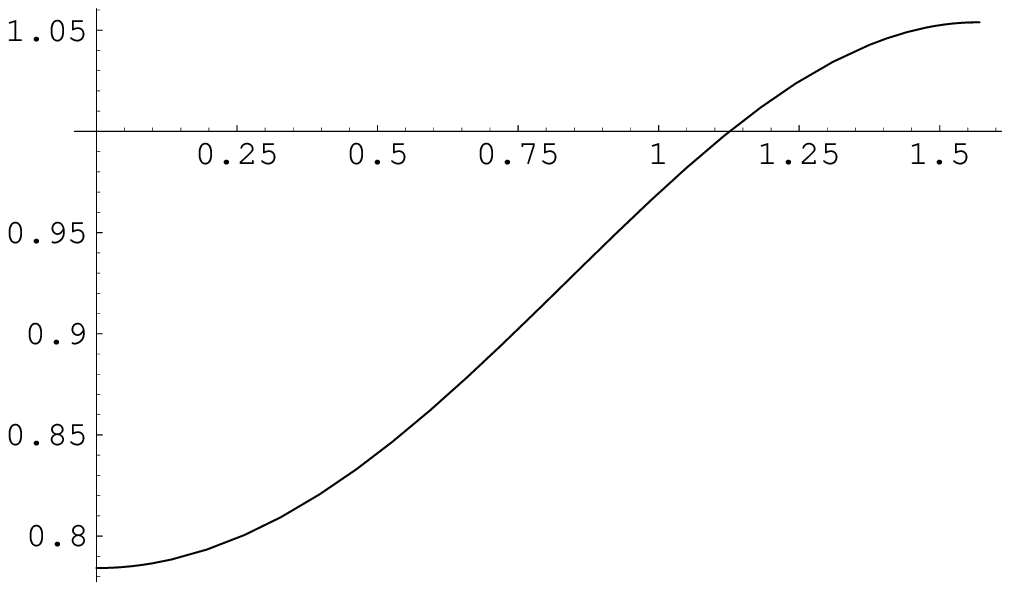}
}
\\
\hspace{-0.0cm} {$\Theta \longrightarrow$ radians}
\caption{The same as in Fig. \ref{fig:kappa1} for a WIMP mass of 100 GeV.}
 \label{fig:kappa2}
  \end{center}
  \end{figure}

%%%%%%%%%%%%%%%%%%%%%%%%%%%%%%%%%%%%%
The parameter $\kappa$ entering the  directional case is shown in Figs. \ref{fig:Skappa1}-\ref{fig:kappa2}.
%%%%%%%%%%%%%%%%%%%%%%%%%%%%%%%%%%%%%%%%%%%%%%%%%%%%%%%%%%%%%%%%%%%%%%%%%%%%%%%%%
%%%%%%%%%%%%%%%%%%%%%%%%%%%%%%%%%%%%%%%%%%%%%%%%%%%%%%%%%%%%%%%%%%%%%%%%%%%%%%%%%

%%%%%%%%%%%%%%%%%%%%%%%%%%%%%%%%%%%%%%%%%%%%%%%%%%%%%%%%%%%%%%%%%%%%%%%%%%%%%%%%%
%%%%%%%%%%%%%%%%%%%%%%%%%%%%%%%%%%%%%%%%%%%%%%%%%%%%%%%%%%%%%%%%%%%%%%%%%%%%%%%%%
%%%%%%%%%%%%%%%%%%%%%%%%%%%%%%%%%%%%%%%%%%%%%%%%%%%%%%%%%%%%%%%%%%%%%%%%%%%%%%%%%

%%%%%%%%%%%%%%%%%%%%%%%%%%%%%%%%%%%%%%
\section{Diurnal Variation}
%%%%%%%%%%%%%%%%%%%%%%%%%%%%%%%%%%%%%
Up to now we have considered the event rate in a directional experiment in fixed direction with respect to the galaxy in the galactic system discussed above. The apparatus, of course, will be oriented in a direction specified in the local frame, e.g. by a point in the sky specified, in the equatorial system, by right ascension $\alpha$ and inclination $\delta$. This will lead to a diurnal variation of the event rate \cite{CYGNUS09}.

The galactic frame, in the so called J2000 system, is defined by the galactic pole with ascension $\alpha_1=12^h~ 51^m~ 26.282^s$ and inclination $\delta_1= +27^0 ~7^{'} ~42.01^{''}$ and the galactic center at $\alpha_2 =17^h ~45^m~ 37.224^s$ , $\delta_2=-(28^0~ 56^{'}~ 10.23^{''} ) $. Thus the galactic unit vector $\hat{y}$, specified by $(\alpha_1,\delta_1)$, and the unit vector $\hat{s}$, specified by $(\alpha_2,\delta_2)$,
can be expressed in terms of the celestial unit vectors $\hat{i}$ (beginning of measuring the right ascension), $\hat{j}$ and $\hat{k}$ (the axis of the Earth's rotation).
One finds
 %\{-0.839866,-0.294454,0.455985\}\{-0.0847317,0.871048,0.483835\}\{0.539652,-0.36772,0.756513\}
 \barr
{\hat y}&=&-0.868{\hat i}- 0.198{\hat j}+0.456{\hat k}
\nonumber \mbox{ (galactic axis) },\\
{\hat x}&=&-{\hat s}=0.055{\hat i}+ 0.873{\hat j}+ 0.483{\hat k} \mbox{ (radially out towards the sun)  },
\nonumber\\
{\hat z}&=&{\hat x}\times{\hat y}=0.494{\hat i}- 0.445{\hat j} +0.747{\hat k}\mbox{ (the sun's direction of motion)}.
\earr
Note in our system the x-axis is opposite to the s-axis used by the astronomers.
Thus a vector oriented by $(\alpha,\delta) $ in the laboratory  is given   in the galactic frame by a unit vector with components:
\beq
\left (
\begin{array}{l}
 y \\
 x \\
 z
\end{array}
\right )
=\left (\begin{array}{l}
 -0.868 \cos {\alpha } \cos {\delta }-0.198 \sin {\alpha
   } \cos {\delta }+0.456 \sin{\delta } \\
 0.055 \cos {\alpha } \cos {\delta }+0.873 \sin {\alpha
   } \cos {\delta }+0.4831 \sin {\delta } \\
 0.494 \cos {\alpha } \cos {\delta }-0.445 \sin {\alpha
   } \cos {\delta }+0.747 \sin {\delta }
\end{array}
\right ).
\eeq
This can also be parameterized \cite{galaxybook} as:
\beq
x=\cos {\gamma } \cos {\delta } \cos \left(\alpha -\alpha _0\right)-\sin
   {\gamma } \left(\frac{}{}\cos {\delta } \cos \left(\theta _P\right) \sin
   \left(\alpha -\alpha _0\right)+\sin {\delta } \sin \left(\theta
   _P\right)\right),
\eeq
\beq
y=\cos \left(\theta _P\right) \sin {\delta }-\cos {\delta } \sin
   \left(\alpha -\alpha _0\right) \sin \left(\theta _P\right),
\eeq
\beq
z=\cos {\delta } \cos \left(\alpha -\alpha _0\right) \sin {\gamma }+\cos
   {\gamma } \left(\frac{}{}\cos {\delta } \cos \left(\theta _P\right) \sin
   \left(\alpha -\alpha _0\right)+\sin {\delta } \sin \left(\theta
   _P\right)\right),
\eeq
where $\alpha_0=282.25^0$ is the right ascension of the equinox, $\gamma\approx 33^0$ was given above and $\theta_P=62.6^0$ is the angle the Earth's north pole forms with the axis of the galaxy. Due to the Earth's rotation this unit vector, with a suitable choice of the initial time, is changing as a function of time
%\beq
%x=\cos \left(\theta _P\right) \sin {\delta }-\cos {\delta } \sin
%   \left(\frac{2 \pi  t}{T}\right) \sin \left(\theta _P\right),
%\eeq
%\beq
\beq
x=\cos {\gamma } \cos {\delta } \cos \left(\frac{2 \pi  t}{T}\right)-\sin
   {\gamma } \left(\frac{}{}\cos {\delta } \cos \left(\theta _P\right) \sin
   \left(\frac{2 \pi  t}{T}\right)+\sin {\delta } \sin \left(\theta
   _P\right)\right),
\eeq

\beq
y=\cos \left(\theta _P\right) \sin {\delta }-\cos {\delta } \sin
   \left(\frac{2 \pi  t}{T}\right) \sin \left(\theta _P\right),
\eeq

\beq
z=\cos \left(\frac{2 \pi  t}{T}\right) \cos {\delta } \sin {\gamma }+\cos
   {\gamma } \left(\cos {\delta } \cos \left(\theta _P\right) \sin
   \left(\frac{2 \pi  t}{T}\right)+\sin {\delta } \sin \left(\theta
   _P\right)\right),
\eeq
where $T$ is the period of the Earth's rotation. In fact the angles $\Theta$ and $\Phi$ discussed above are given by
\beq
\Theta=\cos^{-1}{z},
\eeq
\beq
\Phi=\text{If}\left[x>0,\text{If}\left[y>0,\tan
   ^{-1}\left(\left|\frac{y}{x}\right|\right),\tan
   ^{-1}\left(\left|\frac{y}{x}\right|\right)+\frac{3 \pi
   }{2}\right],\text{If}\left[y>0,\tan
   ^{-1}\left(\left|\frac{y}{x}\right|\right)+\frac{\pi }{2},\tan
   ^{-1}\left(\left|\frac{y}{x}\right|\right)+\pi \right]\right],
\eeq
or
\barr
\Phi&=&H(x)H(y)
\left (\tan ^{-1}\left(\left|\frac{y}{x}\right|\right)\right )+H(x)H(-y)\left (\tan
   ^{-1}\left(\left|\frac{y}{x}\right|\right)+\frac{3 \pi
   }{2}\right )
\nonumber\\
&+&H(-x)H(y)
\left (\tan^{-1}\left(\left|\frac{y}{x}\right|\right)+\frac{\pi }{2}\right )+H(-x)H(-y)
\left (\tan ^{-1}\left(\left|\frac{y}{x}\right|\right)+\pi\right ).
\earr
%\beq
%\Phi=\left %(H(x)H(y)+H(x)H(-y)+3\frac{\pi}{2}+H(-x)H(y)+\frac{\pi}{2}+H(-x)H(-y)+\pi\right)
%\tan  ^{-1}\left(\left|\frac{y}{x}\right|\right)
%\eeq
Some  points of interest are:
\barr
\mbox{The celestial pole: } (y,x,z)&=&(0.460,0.484,0.745)\Rightarrow(\theta=62.6^0,\phi=57^0),
\nonumber\\
\mbox{The ecliptic pole: } (y,x,z)&=&(0.497,0.096,0.863)\Rightarrow(\theta=62.6^0,\phi=83.7^0),
\nonumber\\
\mbox{The equinox: } (y,x,z)&=&(-0.868,0.055,0.494)\Rightarrow(\theta=150.2^0,\phi=83.7^0).
\earr
where $\theta$ is defined with respect to the polar axis (here $y$) and $\phi $ is measured from the $x$ axis towards the $z$ axes.

 The  angle $\Theta$  scanned by the direction of observation is shown, for various inclinations $\delta$, in Fig.~16. We see that for  negative inclinations, the angle $\Theta$ can take values near $\pi$, i.e. opposite to the direction of the sun's velocity, where the rate attains its maximum.     Additionally, the angle $\Phi$ which exhibits rather complicated behavior, is shown for various inclinations $\delta$ in Fig.~17.

%%%%%%%%%%%%%%%%%%%%%%%%%%%%%%%%%%%%%%%%%%%%%%%%%%%%%%%%%%%%%%%%%%%%%%%%%%%%%%%%%
     \begin{figure}[!ht]
 \begin{center}
\rotatebox{90}{\hspace{-0.0cm} {$\Theta  \longrightarrow$ radians}}
     \includegraphics[scale=0.8]{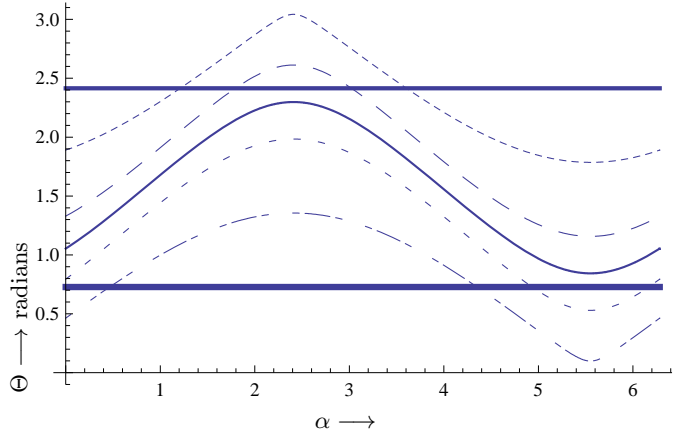}\\
\hspace{-0.0cm} {$\alpha \longrightarrow$}
\caption{ Due to the diurnal motion of the Earth different angles $\Theta$ in galactic coordinates are sampled as the earth rotates. The angle $\Theta$  scanned by the direction of observation is shown for various inclinations $\delta$.  The intermediate thickness, the short dash, the long dash, the fine line, the long-short dash, the short-long-short dash and the thick line correspond to inclination $\delta=-\pi/2,-3\pi/10,-\pi/10,0,\pi/10,3\pi/10$ and $\pi/2$  respectively. We see that, for negative inclinations, the angle $\Theta$ can take values near $\pi$, i.e. opposite to the direction of the sun's velocity, where the rate attains its maximum.}
 \label{fig:theta}
  \end{center}
  \end{figure}
%%%%%%%%%%%%%%%%%%%%%%%%%%%%%%%%%%%%%%%%%%%%%%%%%%%%%%%%%%%%%%%%%%%%%%%%%%%%%%%%%
      \begin{figure}[!ht]
 \begin{center}
\rotatebox{90}{\hspace{-0.0cm} {$\Phi  \longrightarrow$ radians}}
     \includegraphics[scale=0.8]{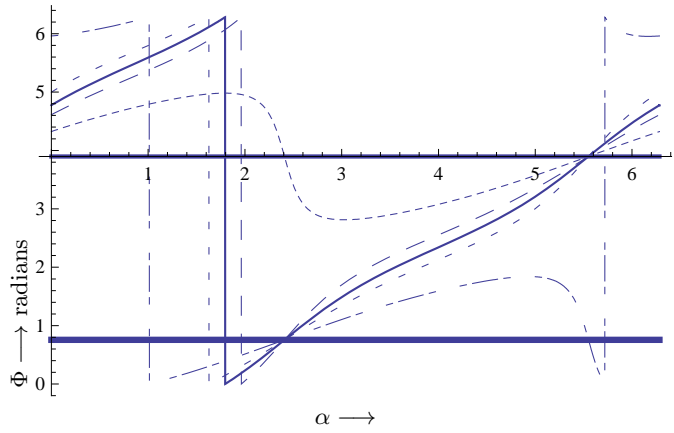}\\
\hspace{-0.0cm} {$\alpha \longrightarrow$}
\caption{The same as in Fig. \ref{fig:theta} in the case of the angle $\Phi$.}
 \label{fig:phi}
  \end{center}
  \end{figure}
\begin{figure}[!ht]
 \begin{center}
     \includegraphics[scale=0.8]{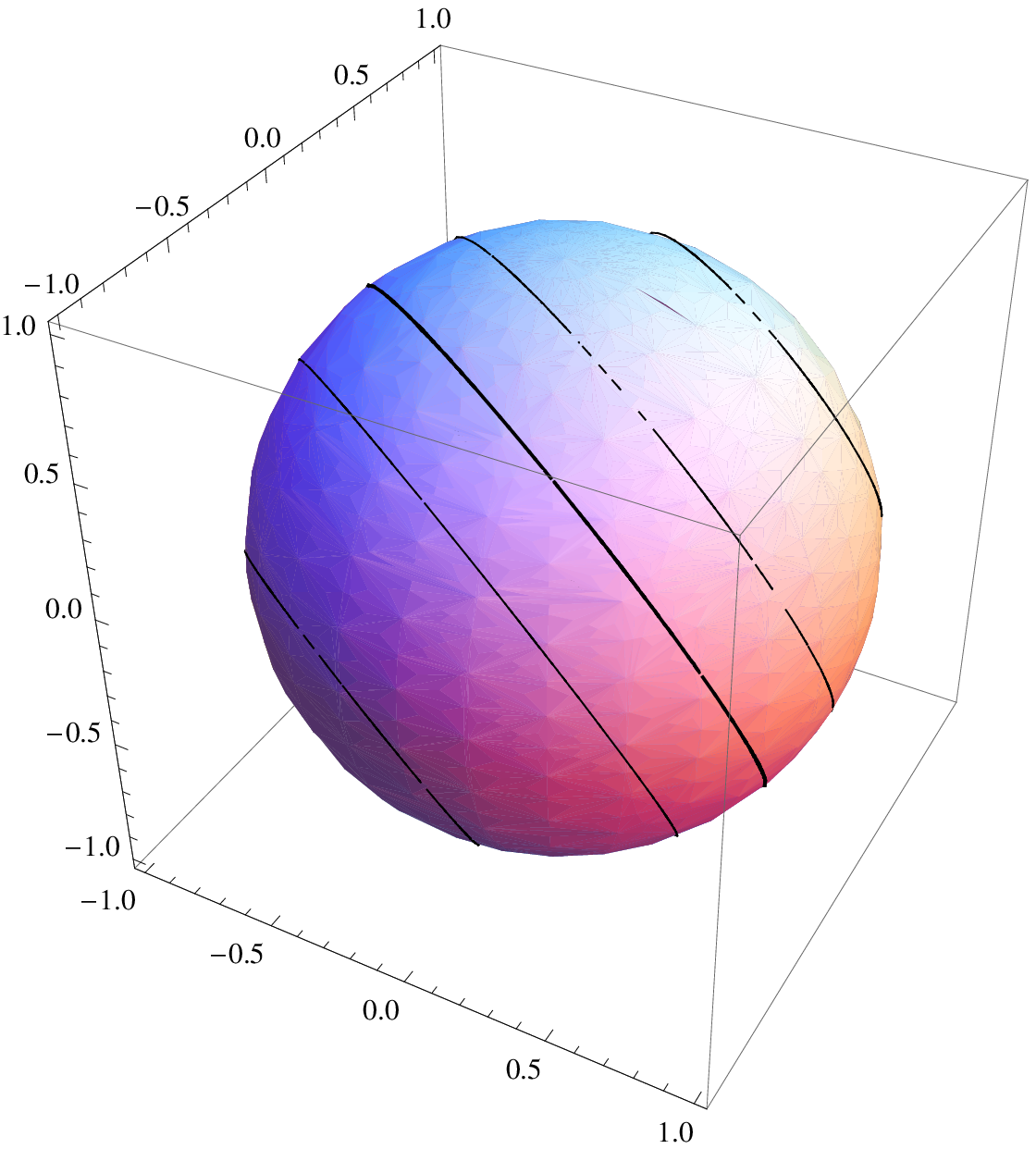}
\caption{The  circular path followed by the point of the direction of observation as seen in the galactic system due to the Earth's rotation for various inclinations $\delta$. The galactic axis is indicated upward. For the path notation see Fig. \ref{fig:theta}.}
 \label{fig:lines}
  \end{center}
  \end{figure}

%%%%%%%%%%%%%%%%%%%%%%%%%%%%%%%%%%%%%%%%%%%%%%%%%%%%%%%%%%%%%%%%%%%%%%%%%%%%%%%%%
 Some typical paths of the orbit of the orientation point in the galactic coordinate system is shown for various inclinations in Fig. \ref{fig:lines}.
%%%%%%%%%%%%%%%%%%%%%%%%%%%%%%%%%%%%%%%%%%%%%%%%%%%%%%%%%%%%%%%%%%%%%%%%%%%%%%%%%
%%%%%%%%%%%%%%%%%%%%%%%%%%%%%%%%%%%%%%%%%%%%%%%%%%%%%%%%%%%%%%%%%%%%%%%%%%%%%%%%%
\\The equipment scans different parts of the galactic sky, i.e. observes different angles $\Theta$. So the rate will change with time depending on whether the sense of the recoiling nucleus can be determined along the line of recoil. The results depend, of course, on the WIMP mass and the target employed. We will consider a light and an intermediate-heavy target.
\subsection{The case of  a light target}
\begin{figure}[!ht]
 \begin{center}
 \subfloat[]
 {
\rotatebox{90}{\hspace{-0.0cm} {sense known $\kappa \longrightarrow$}}
\includegraphics[scale=0.5]{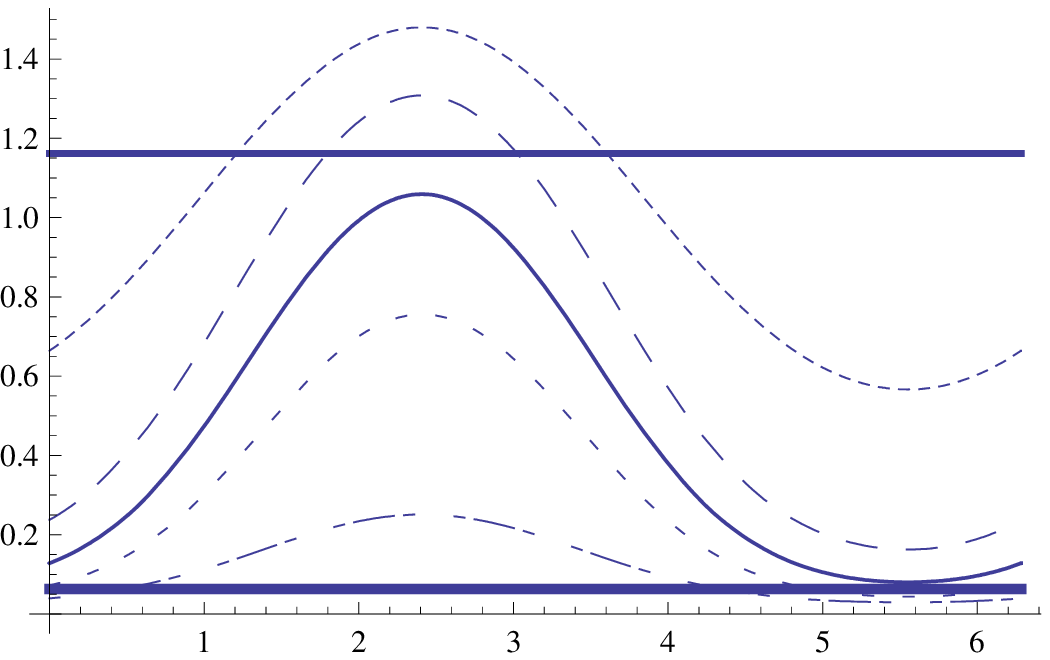}
}
 \hspace{1.0cm}
 \subfloat[]
 {
\rotatebox{90}{\hspace{-0.0cm} {both senses $\kappa \longrightarrow$}}
\includegraphics[scale=0.5]{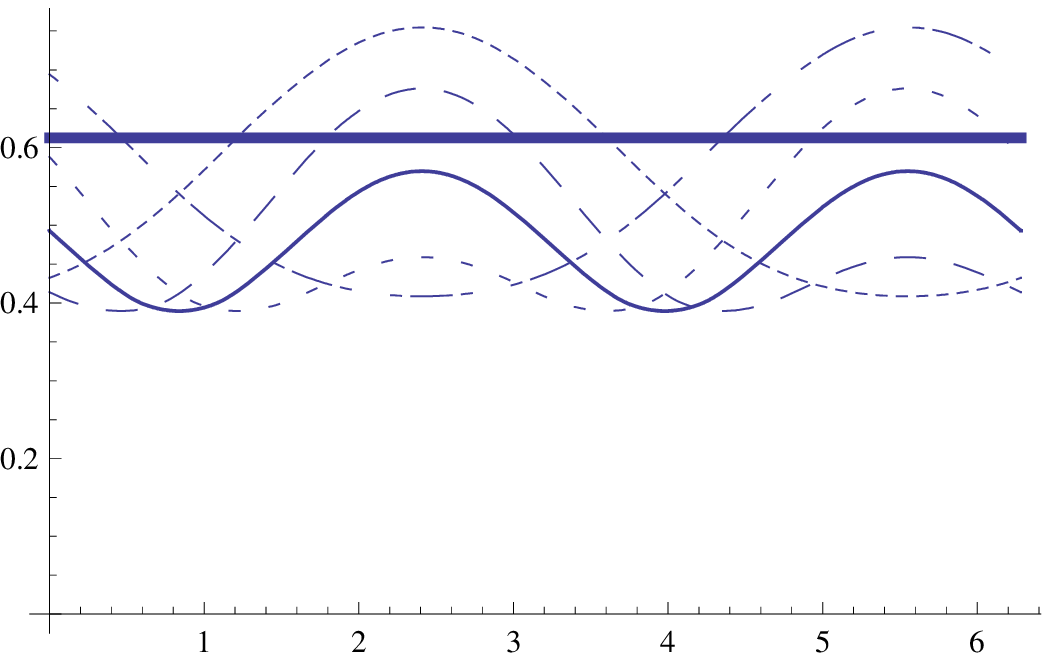}
}
\\
\hspace{-0.0cm} {$t/T \longrightarrow$}
\\
 \subfloat[]
 {
\rotatebox{90}{\hspace{-0.0cm} {sense known $\kappa \longrightarrow$}}
\includegraphics[scale=0.5]{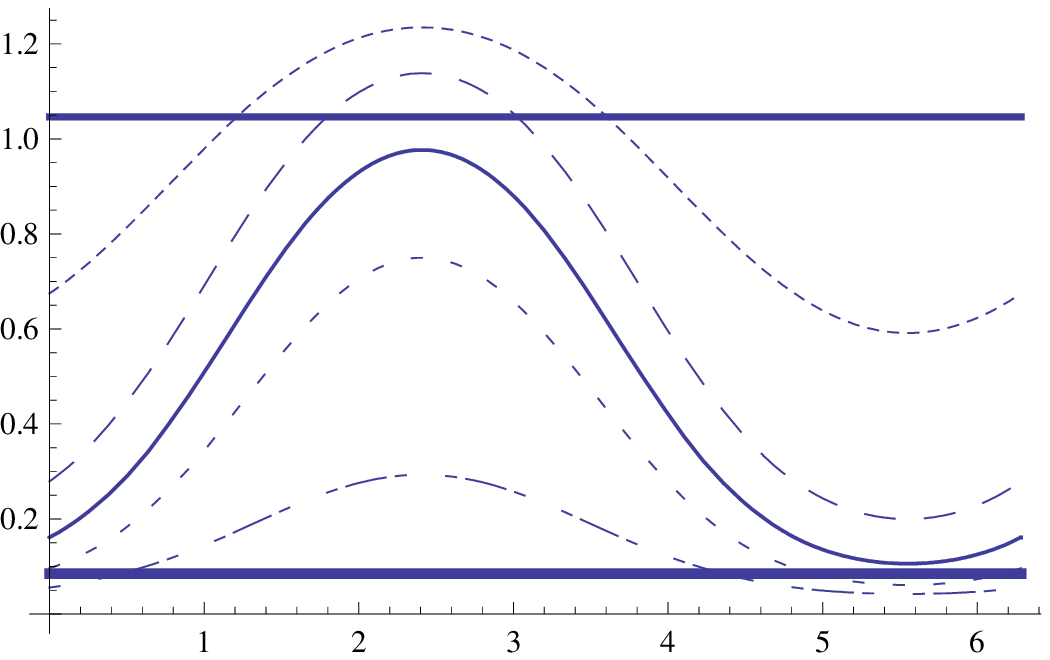}
}
 \hspace{1.0cm}
 \subfloat[]
 {
\rotatebox{90}{\hspace{-0.0cm} {both senses $\kappa \longrightarrow$}}
\includegraphics[scale=0.5]{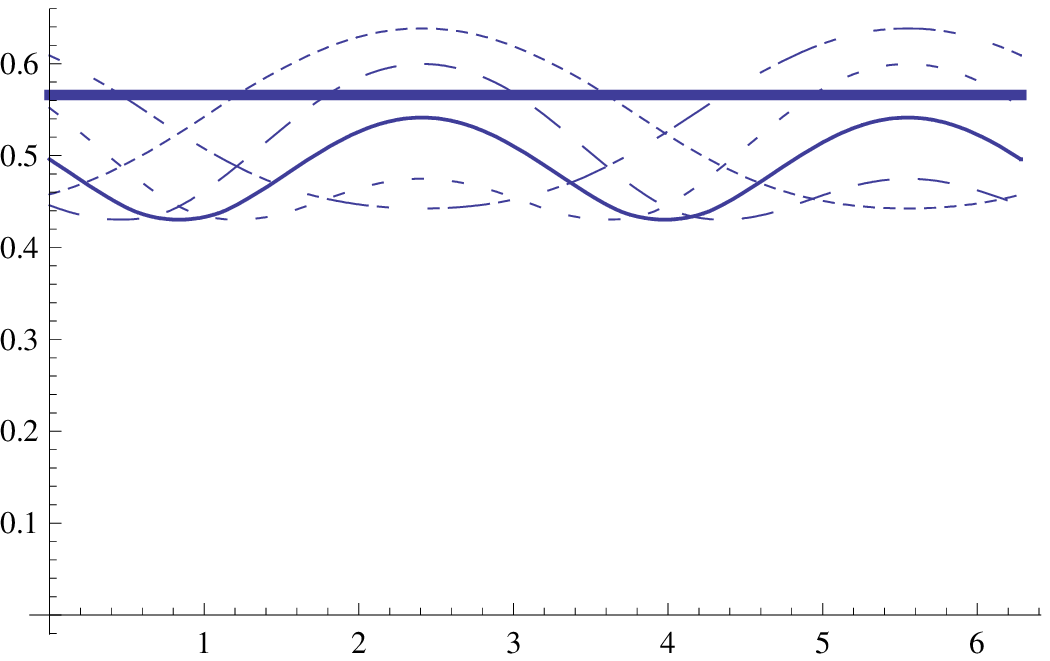}
}
\\
\hspace{-0.0cm} {$t/T \longrightarrow$}
\\
 \subfloat[]
 {
\rotatebox{90}{\hspace{-0.0cm} {sense known $\kappa \longrightarrow$}}
\includegraphics[scale=0.5]{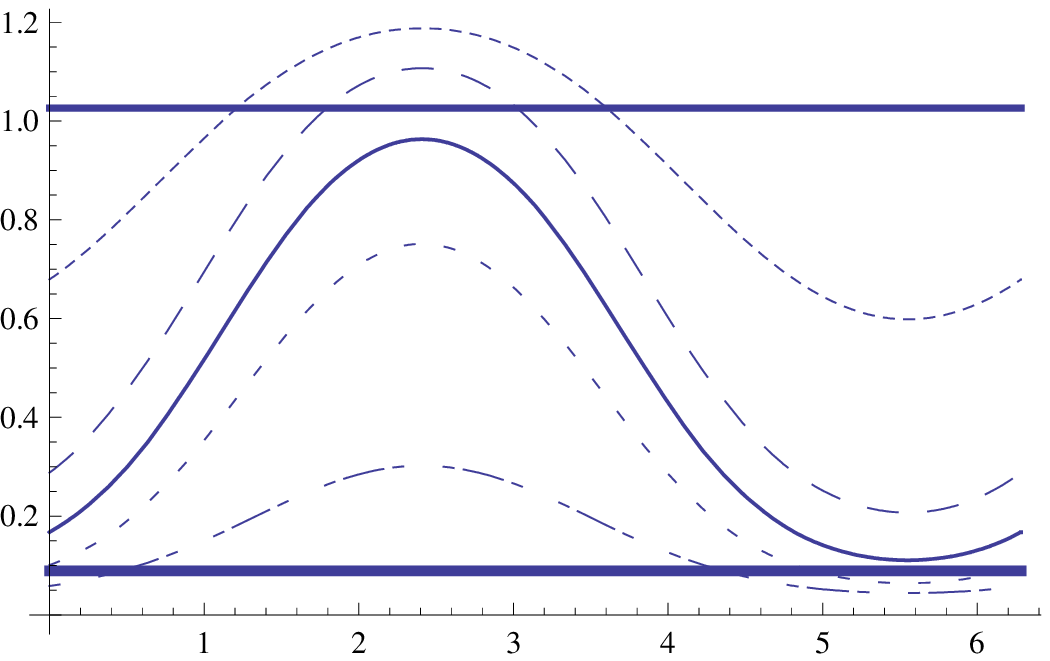}
}
 \hspace{1.0cm}
 \subfloat[]
 {
\rotatebox{90}{\hspace{-0.0cm} {both senses $\kappa \longrightarrow$}}
\includegraphics[scale=0.5]{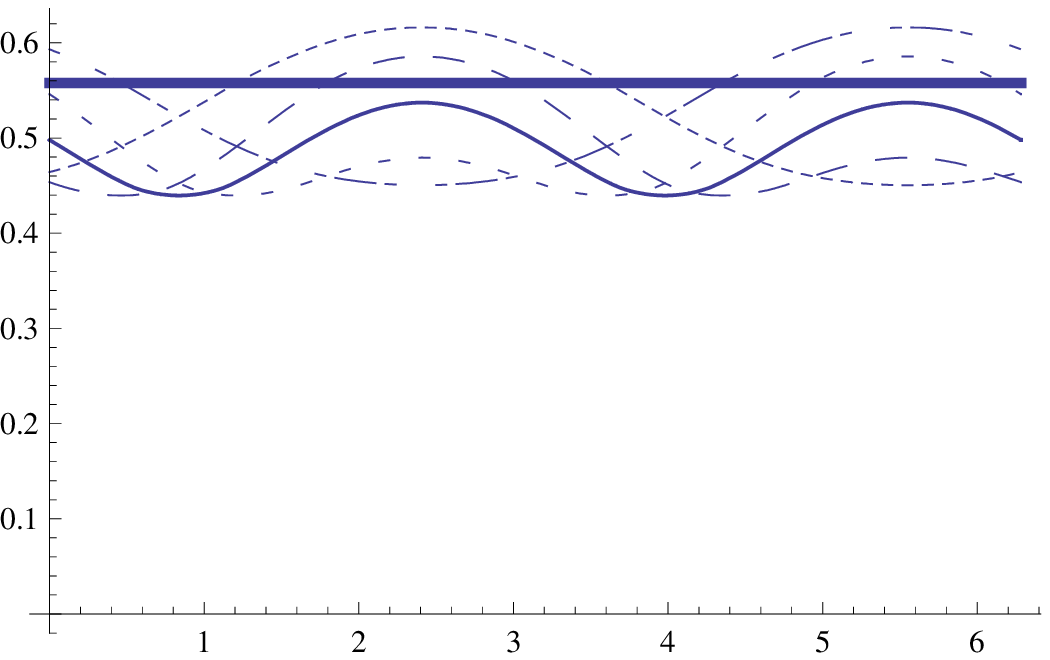}
}\\
\hspace{-0.0cm} {$t/T \longrightarrow$}
\caption{The time dependence(in units of the Earth's rotation period)   of the parameter $\kappa$  for various inclinations $\delta$ in the case of the target CS$_2$.  Due to the diurnal motion of the Earth different angles $\Theta$ are sampled as the earth rotates. For the notation see Figs \ref{fig:kappa2} and \ref{fig:theta}.}
 \label{fig:Sdiurnal}
  \end{center}
  \end{figure}

The time dependence of $\kappa$  is exhibited in Fig. \ref{fig:Sdiurnal}
%%%%%%%%%%%%%%%%%%%%%%%%%%%%%%%%%%%%%%%%%%%%%%%%%%%%%%%%%%%%%%%%%%%%%%%%%%%%%%%%%
%%%%%%%%%%%%%%%%%%%%%%%%%%%%%%%%%%%%%%%%%%%%%%%%%%%%%%%%%%%%%%%%%%%%%%%%%%%%%%%%%
\subsection{The case of  an intermediate-heavy target}
\begin{figure}[!ht]
 \begin{center}
 \subfloat[]
 {
\rotatebox{90}{\hspace{-0.0cm} {sense known $\kappa \longrightarrow$}}
\includegraphics[scale=0.5]{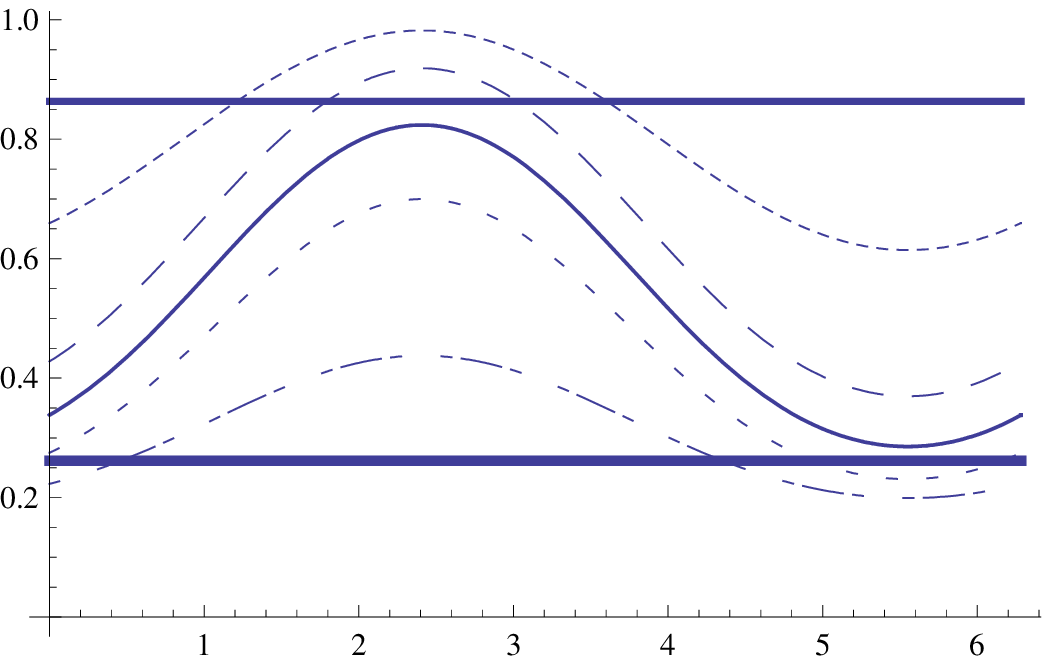}
}
 \hspace{1.0cm}
 \subfloat[]
 {
\rotatebox{90}{\hspace{-0.0cm} {both senses $\kappa \longrightarrow$}}
\includegraphics[scale=0.5]{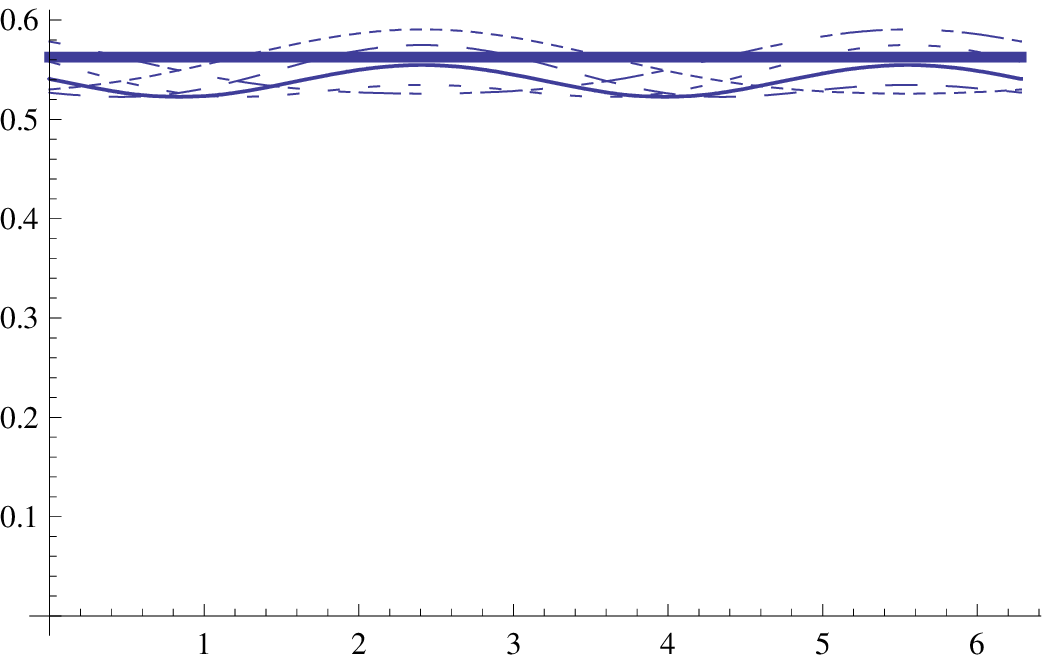}
}
\\
\hspace{-0.0cm} {$t/T \longrightarrow$}
\\
 \subfloat[]
 {
\rotatebox{90}{\hspace{-0.0cm} {sense known $\kappa \longrightarrow$}}
\includegraphics[scale=0.5]{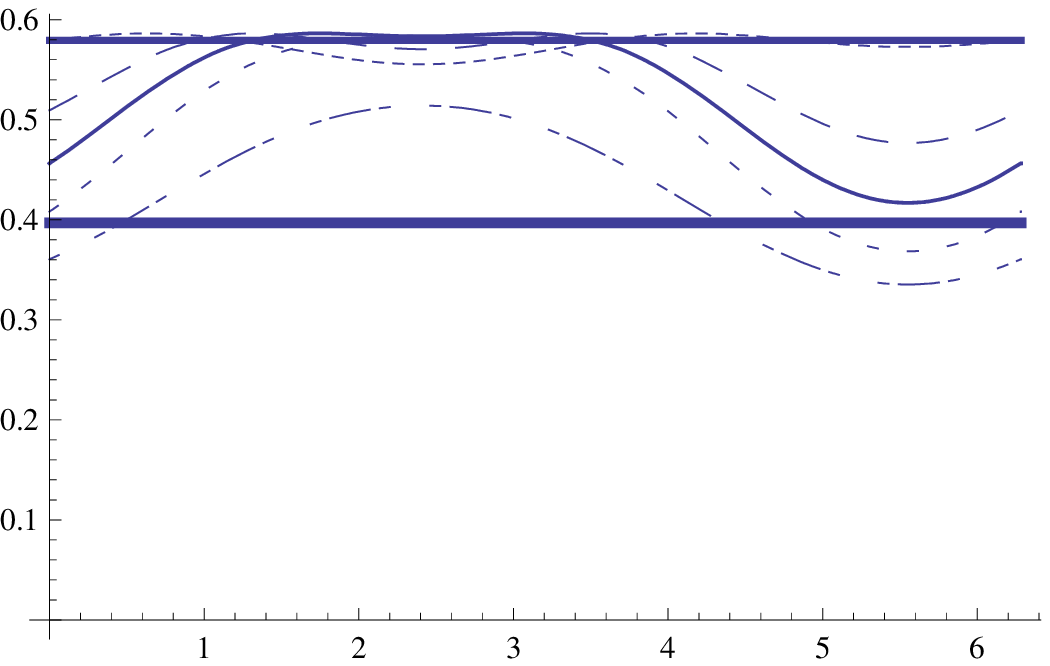}
}
 \hspace{1.0cm}
 \subfloat[]
 {
\rotatebox{90}{\hspace{-0.0cm} {both senses $\kappa \longrightarrow$}}
\includegraphics[scale=0.5]{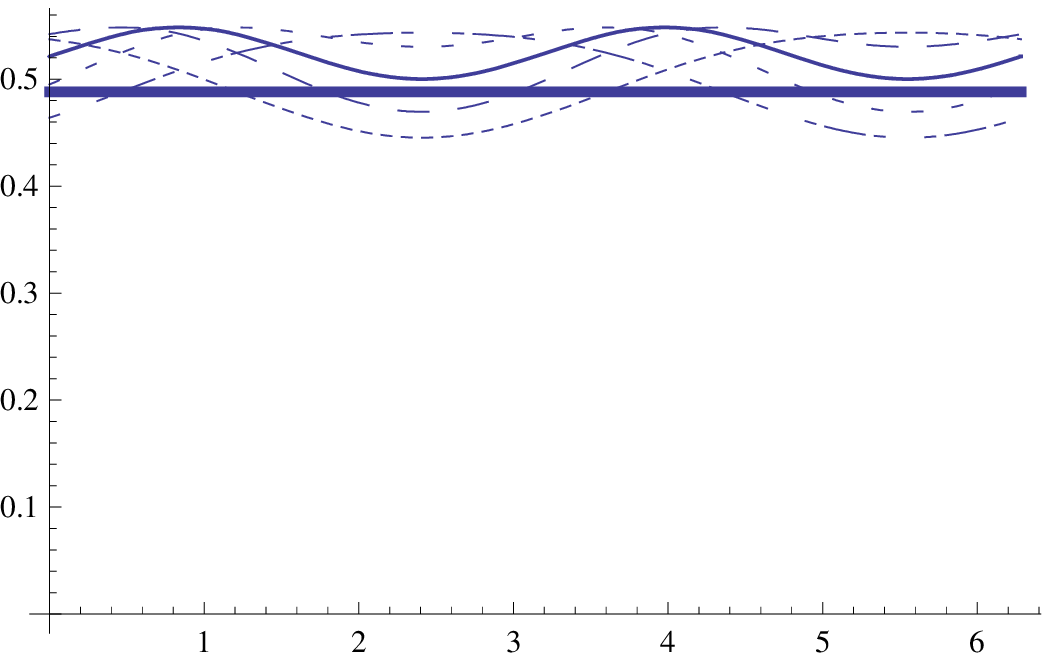}
}
\\
\hspace{-0.0cm} {$t/T \longrightarrow$}
\\
 \subfloat[]
 {
\rotatebox{90}{\hspace{-0.0cm} {sense known $\kappa \longrightarrow$}}
\includegraphics[scale=0.5]{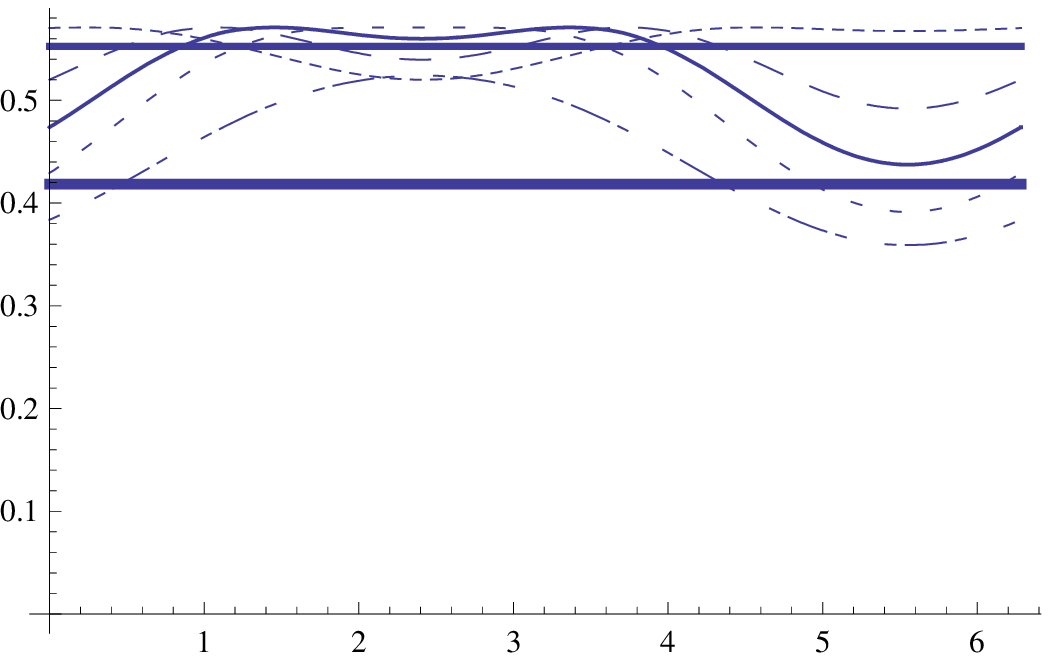}
}
 \hspace{1.0cm}
 \subfloat[]
 {
\rotatebox{90}{\hspace{-0.0cm} {both senses $\kappa \longrightarrow$}}
\includegraphics[scale=0.5]{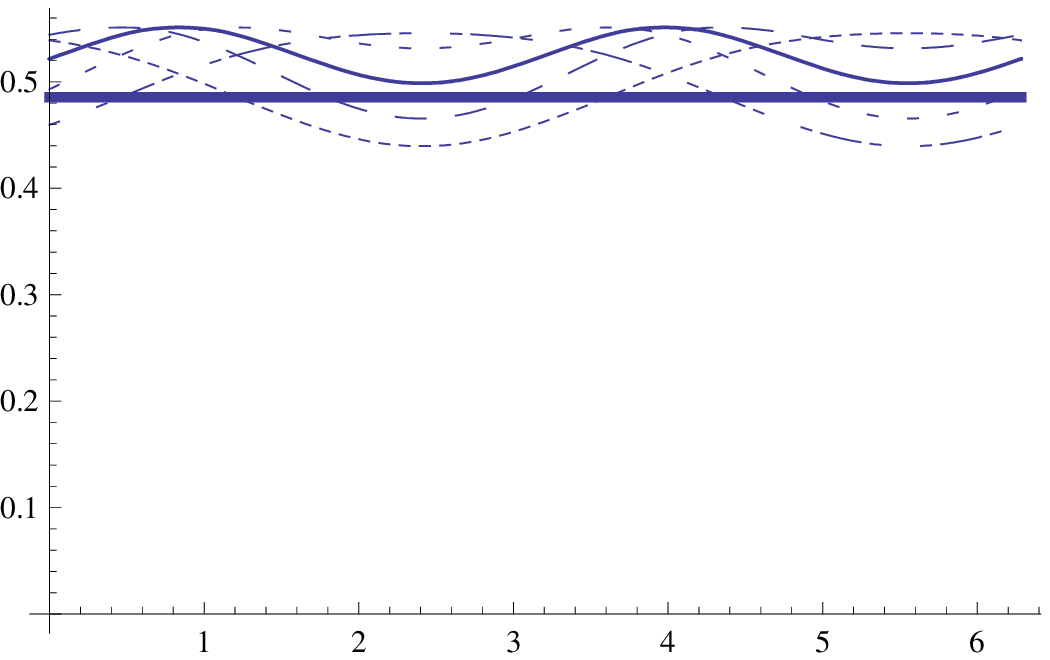}
}\\
\hspace{-0.0cm} {$t/T \longrightarrow$}
\caption{The same  as in Fig. \ref{fig:Sdiurnal} in the case of the Iodine target.}
 \label{fig:Idiurnal}
  \end{center}
  \end{figure}

The time dependence of $\kappa$  is exhibited in Fig \ref{fig:Idiurnal}.
%\ref{fig:rate_diurnal10}-\ref{fig:rate_diurnal200},
%%%%%%%%%%%%%%%%%%%%%%%%%%%%%%%%%%%%%%%%%%%%%%%%%%%%%%%%%%%%%%%%%%%%%%%%%%%%%%%%%

%%%%%%%%%%%%%%%%%%%%%%%%%%%%%%%%%%%%%%%%%%%%%%%%%%%%%%%%%%%%%%%%%%%%%%%%%%%%%%%%%
 \section{Discussion}
 %%%%%%%%%%%%%%%%%%%%%%
 We have seen that, since the expected event rates are small, which means that the background problems become formidable, one should exploit all the signatures provided by the reaction. One such signature is provided by the modulation effect $h$. Unfortunately, however, this effect is small. Furthermore, for heavy targets, its sign, which determines the location of the maximum, is not certain. It depends on the unknown WIMP mass.

 Better signatures are expected in directional experiments in which one measures not only the recoil energy but
 the direction of the nuclear recoil as well.
  Some of the requirements that should be met by such detectors have recently been discussed \cite{CKS-DS05,GREEN06}.
 To fully exploit the advantages of such detectors one should be able to distinguish between recoils with
momenta ${\bf p}$ and $-{\bf p}$ (sense of direction), which now appears to be feasible.
 Some of the predicted interesting features of directional event rates persist, even if it turns out
that the sense of motion of recoils along their line of motion cannot be measured.

Such experiments given a sufficient number of events provide an excellent signature to discriminate against background.
One, of course, gets a smaller rate by observing  in a given direction. In the most favored direction, opposite to the sun's direction of
motion, the event rate is $\approx \frac{\kappa}{2 \pi}$ down from that of the
standard non directional experiments, if a specific angle $\Phi$ is chosen.  Since, however, $\kappa$ is independent of the angle $\Phi$, one can integrate over all azymouthal  and thus the retardation is only $\kappa\approx1$.

Finally we have seen that in directional experiments the relative event rate for detecting WIMPs within our galaxy, as given by the parameter $\kappa$, will show a periodic diurnal variation due to the rotation of the Earth. The parameter $\kappa$ is essentially independent of any particle model parameters other than the WIMP mass. It does depend on the assumed velocity distribution.  The time variation is larger in the case of light WIMP and/or light target. In all cases it is preferable to select a direction of observation  with negative inclination, since, then,  both the absolute value of $\kappa$ and its variation are larger.
The time variation is larger in fully directional experiments. But, even if  the sense along the line of motion is not observed, one will observe a time variation, but it will be smaller. This is hardly surprising, however, since, then, the observed event rate is the average of the two senses of direction. Note that in this case the time required to come to the same point is half of that of  the previous case (see Figs \ref{fig:Sdiurnal}-\ref{fig:Idiurnal}).

 The time variation arising after the inclusion of the modulation parameters ($h_m, \alpha$) or ($h_c,h_s$) is expected to be even more complicated, since these parameters depend on both $\Theta$ and $\Phi$. One expects a diurnal variation on top of the annual variation characteristic of the usual modulation effect entering both directional and non directional experiments. Such effects will be discussed elsewhere.
      \section*{Acknowledgments} The work of one of the authors (JDV) was partially supported by the  grant MRTN-CT-2006-035863 (UniverseNet). He is indebted to Professors N. Spooner and J. Jochum for useful discussions
on experimental aspects of this work.
%\include{inserthm}
%\include{insertalpham}
%\bibliography{TeX}
%\end{document}

\end{document}